\documentclass[12pt]{iopart}

\usepackage{bm}
\usepackage{bbm}
\usepackage{amssymb}
\usepackage{braket}
\usepackage{graphicx}
\usepackage{xcolor}
\usepackage[normalem]{ulem}

\newcommand{\SubFig}[2]{\ref{#1}{\color{blue}#2}}
\newcommand{\ketbra}[2]{\ket{#1}\!\bra{#2}}
\newcommand{\eqref}[1]{(\ref{#1})}
\newcommand{\normord}[1]{:\!#1\!:}

\newcommand{\openone}{\mathbbm{1}}

\usepackage{orcidlink}

\begin{document}

\title[Time-delayed collective dynamics in waveguide QED and bosonic quantum networks]{Time-delayed collective dynamics in waveguide QED and bosonic quantum networks}

\author{Carlos Barahona-Pascual\orcidlink{0009-0007-2372-9397}}
\address{Instituto de Física Fundamental, IFF-CSIC, Calle Serrano 113b, Madrid 28006}

\author{Hong Jiang\orcidlink{0009-0008-7937-6671}}
\address{Instituto de Física Fundamental, IFF-CSIC, Calle Serrano 113b, Madrid 28006}

\author{Alan C. Santos\orcidlink{0000-0002-6989-7958}}
\address{Instituto de Física Fundamental, IFF-CSIC, Calle Serrano 113b, Madrid 28006}
\ead{ac\_santos@iff.csic.es}

\author{Juan José García-Ripoll\orcidlink{0000-0001-8993-4624}}
\address{Instituto de Física Fundamental, IFF-CSIC, Calle Serrano 113b, Madrid 28006}
\ead{jj.garcia.ripoll@csic.es}

\begin{abstract}
This work introduces a theoretical framework to model the collective dynamics of quantum emitters in highly non-Markovian environments, interacting through the exchange of photons with significant retardations. The formalism consists on a set of coupled delay differential equations for the emitter's raising/lowering operators $\sigma^\pm_i$, supplemented by input-output relations that describe the field mediating the interactions. These equations capture the dynamics of both linear (bosonic) and nonlinear (two-level) emitter arrays. It is exact in some limits---e.g., bosonic emitters or generic systems with up to one collective excitation---and can be integrated to provide accurate results for larger numbers of photons. These equations support a study of collective spontaneous emission of emitter arrays in open waveguide-QED environments. This study uncovers an effect we term cascaded super- and sub-radiance, characterized by light-cone-limited propagation and increasingly correlated photon emission across distant emitters. The collective nature of this dynamics for two-level systems is evident both in the enhancement of collective emission rates, as well as in a superradiant burst with a faster than linear growth. While these effects should be observable in existing circuit QED devices or slight generalizations thereof, the formalism put forward in this work can be extended to model other systems such as network of quantum emitters or the generation of correlated photon states.
\end{abstract}
\newpage 

\section{Introduction}
\label{sec:introduction}
The study of light-matter interacting systems offers unique opportunities to discover emerging collective phenomena such as the super-radiant emission of light~\cite{dicke1954, gross1982}, and its counterpart, the sub-radiant regime~\cite{pavolini1985, devoe1996}. In particular, setups of emitters coupled to quasi-1D photonic environments---known as waveguide quantum electrodynamics or waveguide-QED~\cite{sheremet2023}---facilitate the quest for collective phenomena thanks to (i) strong light-matter interactions, (ii) novel long-range coherent and incoherente interactions mediated by long-lived propagating photons and (iii) a huge variety of potential implementations, such as atoms coupled to waveguides~\cite{solano2017, liedl2024} and photonic crystals~\cite{goban2015}, quantum dots in photonic nanostructures~\cite{lodahl2015}, superconducting circuits~\cite{forn-diaz2017,wang2020a} and matter-wave quantum emitters~\cite{krinner2018}.

So far, the study of collective phenomena in light-matter systems has been dominated by a limit in which emitters are nearby packed and the photonic environment acts as an infinite, Markovian bath, which supports instantaneous coherent and incoherent interactions~\cite{gonzalez-tudela2011, shahmoon2013}. However, recent waveguide-QED setups have opened the door to interactions mediated by non-Markovian environments. Prominent examples include long-distance quantum links among superconducting quantum processors~\cite{zhong2018, leung2019, magnard2020, chang2020, zhong2021, storz2023, niu2023, grebel2024, qiu2025}, experiments with artificial atoms in slow-light waveguides~\cite{ferreira2024}, giant atoms in hybrid superconductor-SAW devices~\cite{ferreira2024,gustafsson2014,andersson2019}, and with matter-wave quantum emitters~\cite{kim2025}.

The description and modelization of strongly non-Markovian systems with significant delays, unbounded memory and strong interactions is a very relevant theoretical challenge. In scenarios in which the environment is well known, this problem has been addressed using a combination of analytical,  semi-analytical and purely numerical treatments. For one or two excitations, a Wigner-Weisskopf model~\cite{diaz-camacho2015, sinha2020, sinha2020a, qiu2023, Gu2024,Lanuza2022,lanuza2024} reduces to a manageable set of coupled delayed differential equations. For more photons, particular setups with chiral couplings~\cite{liedl2024} or which cancel the backward influence of propagating photons~\cite{magnard2020}, can be modeled as cascaded systems with sequences of physical transformations~\cite{kiilerich2019} and integrable master equations~\cite{cirac1997, windt2025a}. These methodologies have been supplemented by a large family of numerical methods based on tensor network formalisms, such as the TEMPO algorithm~\cite{strathearn2018}, numerical renormalization techniques with multiple emitters in 1D, 2D and 3D environments~\cite{feiguin2020}, and a discretization of time-delayed dynamics and quantum feedback using tensor networks~\cite{vodenkova2024, pichler2016}.

As an alternative solution to this long-standing problem, this work introduces a theoretical framework for bosonic quantum networks of emitters interacting strongly by the exchange of photons with significant retardations. The framework relies on a Heisenberg-Langevin (HL) formalism to model the emitters with a closed set of delay-differential operator equations. These are supplemented with input-output relations that account for the propagating fields mediating the interactions. The HL equations are exact in some limits and can be numerically integrated, for any number of photons, up to a moderate number of emitters $N\sim 10$, providing results that agree with exact two-photon simulations---up to what can be expected from finite-size problems---in simulations that scale more favorably than previously developed methods.

The discovery of these equations allows us to explore new physics in waveguide-QED setups. In particular, we apply the new technique to study super- and subradiant emission in arrays of emitters that are coupled to a 1D infinite waveguide and experience strongly retarded, photon-mediated interactions. By working both with linear and saturable emitters, we provide evidence for the emergence of collective phenomena in the saturable regime. More precisely, we demonstrate the appearance of a \textit{cascaded superradiant} behavior in which emitters get progressively correlated by the arrival from photons from further away neighbors. This phenomenon manifests in the time-dependent growth of the collective spontaneous emission rate and in the emergence of a superradiant peak with a superlinear growth $(N^\alpha,\; 1<\alpha \leq 2)$ that is markedly different from the Markovian regime and which depends on the emitters separation. These phenomena may be observed in waveguide-QED microwave networks~\cite{magnard2020, zhong2021, qiu2025}, or similar setups on chip~\cite{zhong2018, ferreira2024}.

\begin{figure}[t]
	\centering
	\includegraphics[width=0.55\columnwidth]{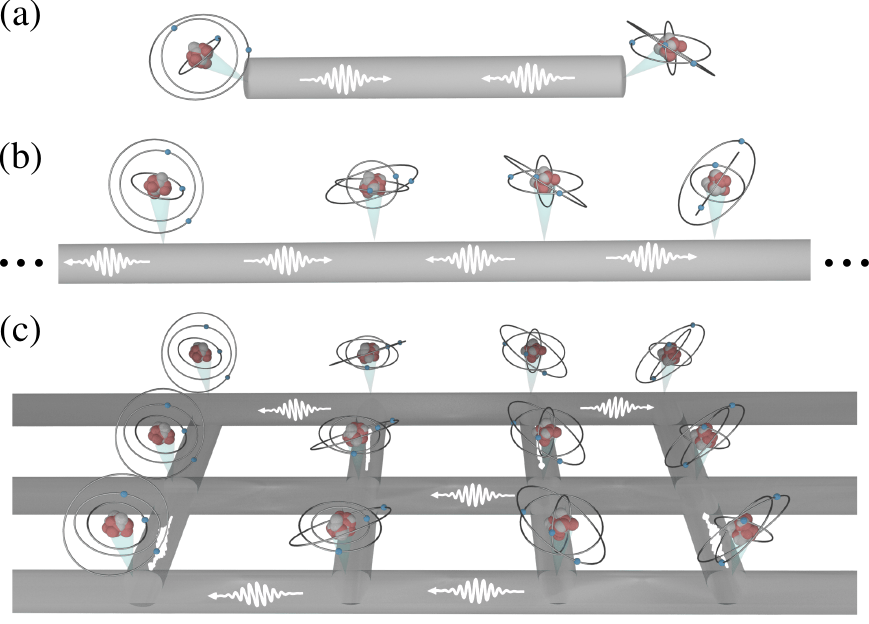}
	\caption{Quantum emitters (atoms, for instance) connected to a non-Markovian network (blue shades). The emitters interact by exchanging photons that travel through the common bosonic network or bath. The delays of the photons $\tau_{ij}\propto d_{ij}/c$ when travelling between any pair of emtitters $i$ and $j$, causes the network to act as a non-Markovian environment. In this work we mainly explore (a) open and (b) closed one-dimensional environments, but the formalism applies to more general bosonic quantum networks (c).}
	\label{fig:setup}
\end{figure}

The outline of this manuscript is as follows. Sect.~\ref{sec:theory} presents a \textit{bosonic quantum network} of emitters connected by one-dimensional photonic environments. An HL formalism produces exact integro-differential equations for the quantum emitters. These become simpler delay-differential equations (DDEs) when the emitters are linear and, when the emitters are saturable, through a new approximation (c.f. Sect.~\ref{sec:dde-tls}). Sect.~\ref{sec:applications} demonstrates that the DDEs are exact in some limits, and derive specific expressions for setups with emitter arrays connected to open waveguides and for a quantum link. Sect.~\ref{sec:validation} provides evidence that these equations provide quantitatively good results for problems with up to two excitations, with deviations that can be explained by the same finite-size effects that make a single emitter deviate from a purely exponential decay. With this in mind, Sect.~\ref{sec:cascaded} explores new physics of spontaneous emission from emitter arrays, demonstrating the emergence of collective phenomena and the \textit{cascaded superradiance} described before. Finally, in Sect.~\ref{sec:discussion} we close with a discussion of the main results and open lines of research and applications in the near future.

\section{Non-Markovian theory of quantum emitters in bosonic waveguide}%
\label{sec:theory}
\subsection{Bosonic quantum network}
\label{sec:model}
In this work, we aim to study networks of quantum emitters that exchange information via photons that travel in a low-dimensional bosonic, such as a one-dimensional waveguide or a network spanning multiple nodes in a more complex topology (c.f. Fig.~\ref{fig:setup}). The model we use to describe this system is a Hamiltonian defined in the composite Hilbert space of the quantum nodes and the bosonic network $\mathcal{H}_{\mathrm{E}} \otimes \mathcal{H}_{\mathrm{WG}}$
\begin{eqnarray}
	\label{eq:the-model}
	\hat{H}
	 &=& \hat{H}_e + \hat{H}_{wg} + \hat{H}_I\\
	 &=&  \sum\nolimits_{n = 1}^{N} \Delta_{n} \hat{s}_{n}^{\dagger}\hat{s}_{n}  +
	\sum\nolimits_{\mu}\int\nolimits_{0}^{\infty}  \omega  \hat{a}_{\omega,\mu}^{\dagger}\hat{a}_{\omega,\mu} d \omega
	\nonumber \\
	&=&+  \sum_{\mu} \sum_{n=1}^{N}\int_{0}^{\infty}  \left(V_{n,\mu}^{\ast}(\omega) \hat{a}_{\omega,\mu}^{\dagger}\hat{s}_{n} + V_{n,\mu} (\omega)  \hat{s}_{n}^{\dagger}\hat{a}_{\omega,\mu}\right)d\omega.
\end{eqnarray}
In this model, the emitters are bosonic objects with raising and lowering operators  $\hat{s}_{n}$ and $\hat{s}_{n}^{\dagger}$, to be replaced with Fock operators $\{\hat{b}_m,\hat{b}_m^\dagger\}$ such that $[\hat{b}_m,\hat{b}_m^\dagger]=1$ if the nodes are actual cavities or resonators, or with Pauli matrices $\hat{\sigma}^-_m=\ketbra{0}{1}$ and $\hat{\sigma}^+_m=\ketbra{1}{0}$ if the emitters are two-level systems.

The emitters interact with a collection of photonic modes labeled both by the frequency and by some other quantum numbers. For instance, in the one-dimensional infinite waveguide from Fig.~\SubFig{fig:setup}{b} the modes can be right- or left-ward moving, $\hat{a}_{\omega,\rightarrow}, \hat{a}_{\omega,\leftarrow}$. In the open waveguide from Fig.~\SubFig{fig:setup}{a} this label disappears, as we work with standing waves. And in more general networks, other labels may be required. In all these scenarios, we assume that the photons traverse the cables or waveguides at a finite group velocity, $v_\mu$, creating the conditions for a non-Markovian dynamic. More precisely, we ask that the delays $|d_{ij}/v_\mu|$ experienced by photons travelling between any pair of emitters $i$ and $j$ (c.f. Fig.~\SubFig{fig:setup}{a}) are comparable or greater than the length of the photons, giving rise to \textit{an effective environment with long-term memory}.

As a final technical detail, we assume a Jaynes-Cummings type interaction between emitters and photons. This number-conserving model arises from a rotating wave approximation that is correct even in the strong coupling limit, provided the typical spontaneous emission rates are slower than the frequencies $\Delta_n$ of the photons created in the network. The emitter-waveguide interaction profile $V_{n,\mu}(\omega)$ is assumed to be a broad and smoothly varying function in frequency space. For example, by considering that each emitter interacts with the modes of the waveguide in the dipolar approximation, we can write $V_{n,\mu}(\omega) = g_{n,\mu}(\omega) e^{i x_{n} \omega / v_{\mu}}$, where $x_{n}$ is the position of the $n$-th emitter along the waveguide. Interactions with chiral nature and with other complex emitters can be taken into account, but some non-local objects such as giant atoms~\cite{kannan2020} introduce subtleties in the treatment below that need to be separately addressed.

\subsection{The Heisenberg-Langevin equations}%
\label{sec:genericHL}

Ideally, the emitters and the photonic network form a closed system, whose dynamics is described by the Schrödinger equation, with a unitary operator $\hat{U}(t)$ satisfying $i\partial_t \hat{U}(t)=H\hat{U}(t)$, that provides us with the quantum state of the system at any other time obtained from $\ket{\psi(t)} = \hat{U}(t)\ket{\psi(0)}$, as well as with predictions about the expected values of any (time-independent) observable $\hat{o}$, given by $\braket{\hat{o}}_t = \braket{ \psi(t)| \hat{o}|\psi(t)}$.

In this work we will instead use the Heisenberg picture, where observables evolve in time according to the transformation $ \hat{o}(t) := U(t)^\dagger\hat{o}U(t)$. In this picture, expectation values are recovered using the initial states $\braket{\hat{o}(t)} := \braket{\psi(0)|\hat{o}(t)|\psi(0)}$ and the observables follow a dynamical equation generated by the Hamiltonian
\begin{equation}
	\label{eq:Heisenberg}
	\frac{d}{dt}\hat{o}(t) = -i [\hat{o}(t), \hat{H}],\; \mbox{with } \hat{o}(0) =\hat{o}.
\end{equation}

Four important remarks follow. First, note that the Hamiltonian $\hat{H}$ is an invariant of motion, but must be expressed using the time-evolved versions of the operators---i.e., $\hat{a}$, $\hat{s}$, $\hat{a}^\dagger$, etc. are replaced with the time-dependent versions. Second, commutation relations are preserved by the evolution---e.g., $[\hat{a}_{k}(t),\hat{a}_{k'}^\dagger(t)]=\delta_{kk'}$ and $[\hat{s}_{l}^{\dagger}(t),\hat{s}_{m}(t)]=[\hat{s}_{l}^{\dagger},\hat{s}_{m}]$. This makes the actual computation of the Eq.~\eqref{eq:Heisenberg} feasible. Third, Eq.~\eqref{eq:Heisenberg} has been particularized for observables $\hat{o}$ that are time-independent. If we were interested in operators such as $\hat{o}_t = \cos(\omega t)\hat{s}+ \sin(\omega t)\hat{s}^\dagger$, the equation would acquire an additional $\partial_t \hat{o}_t$. Finally, as it is usual in quantum optics, we will distinguish the Heisenberg picture by labelling the operators with their time dependence $\hat{o}(t)$, unless we may unambiguously drop the label.

The HL equations for the emitter and the field as described by the model~\eqref{eq:the-model} read
\begin{eqnarray}
	\frac{d\hat{s}_{l}(t)}{dt} &= i\hat{s}^{0}_{l}(t) \left[\Delta_{l}\hat{s}_{l}(t)  +  \sum_{\mu = \rightleftarrows}\int\nolimits_{0}^{\infty} V_{l,\mu}(\omega)\hat{a}_{\omega,\mu}(t) d\omega\right] \nonumber \\
	\frac{d\hat{a}_{\omega,\mu}(t)}{dt} &= -i\omega_{k}\hat{a}_{\omega,\mu}(t)-i \sum\nolimits_{j=1}^{N}V_{j,\mu}^{\ast}(\omega)\hat{s}_{j}(t). \label{HL_Equations}
\end{eqnarray}
The operator $\hat{s}^{0}_{l}(t) = [\hat{s}_{l}^{\dagger}(t),\hat{s}_{l}(t)]$ will be $-\openone$ for bosons and $\hat{\sigma}^z$ for qubits. The dynamics of the bosonic environment may be integrated formally
\begin{equation}
	\hat{a}_{\omega,\mu}(t)= \hat{a}_{\omega,\mu}(0)e^{-i\omega t}-i\int_{0}^{t}\sum_{j=1}^{N}V_{j,\mu}^{\ast}(\omega)e^{-i\omega(t-\tau)}\hat{s}_{j}(\tau) d\tau ,
	\label{eq:field}
\end{equation}
and replaced into the emitter's equation, to create a stand-alone integro-differential equation. By performing an interaction-like picture transformation $\hat{s}_{l}(t) \rightarrow \hat{s}_{l}(t)e^{-i\Delta_{l}t}$, this equation simplifies to
\begin{equation}
	\label{eq:integro_dif}
	\frac{d\hat{s}_{l}^{-}(t)}{dt} = \hat{s}^{0}_{l}(t) \left[\hat{\xi}_{l} (t) +  \sum_{j=1}^{N} \int_{0}^{t} K_{lj}(t-\tau) e^{i(\Delta_{l}t-\Delta_{j}\tau)}\hat{s}_{j}(\tau) d \tau\right].
\end{equation}
The input electromagnetic field at the $l$-th emitter
\begin{equation}
	\hat{\xi}_{l}(t) = i\sum_{\mu = \rightleftarrows}\int\nolimits_{0}^{\infty} V_{l,\mu}^{\ast}(\omega)\hat{a}_{\omega,\mu}(0)e^{-i(\omega-\Delta_{l}) t} d\omega,
\end{equation}
introduces the field fluctuations from the waveguide's initial state via the time-independent operators $\hat{a}_{\omega,\mu}$, and the memory function kernel
\begin{equation}
	\label{eq:memory-function}
	K_{lj} (t-\tau)= \sum_{\mu = \rightleftarrows}\int\nolimits_{0}^{\infty}V_{l,\mu}(\omega)V_{j,\mu}^{\ast}(\omega)e^{-i\omega(t-\tau)} d\omega.
\end{equation}
describes the emitters' decay ($l=j$) and the waveguide-mediated coherent and dissipative interactions ($l\neq j$). Note that this second term reflects the memory of the environment, which is not Markovian, and keeps track of correlations of the emitters with the stream of photons that traverses the photonic network.

Equations~\eqref{eq:integro_dif}-\eqref{eq:memory-function} are the central tool in this work. First, by analyzing the memory functions~\eqref{eq:memory-function}, we will be able to transform the integro-differential equation into a set of approximate delay-differential equations that describe the collective dynamics of the emitters in a regime of dilute photons. Second, this equation will be benchmarked and shown to describe accurately models that range from collective spontaneous emission of ensembles in free space, all the way to quantum state transfer between nodes in a quantum link. Finally, we will demonstrate how the explicit form of the delay-differential equations provides us with a quantitative and qualitative description of superradiance physics, introducing a new phenomenon that we call cascaded superradiance, and which can be fully explained with the delayed memory functions from Eq.~\eqref{eq:integro_dif}.

\subsection{Time-delayed interactions for linear emitters}
\label{sec:dde-cavity}
As a first application of the previous theory, let us discuss a network of linear emitters, such as a set of quantum harmonic oscillators interconnected by the photonic waveguides~\cite{xiang2017, xiang2023}. Eq.~\eqref{eq:integro_dif} simplifies to
\begin{equation}
	\frac{d}{dt} \hat{s}_{l}(t) = - \hat{\xi}_l(t) - \sum_{j=1}^{N} \int_{0}^{t} K_{lj}(t-\tau)e^{i\Delta(t-\tau)} \hat{s}_{j}(\tau) d \tau, \label{Eq:Bcavity}
\end{equation}
with the bosonic operators $[\hat{s}_l,\hat{s}_j^\dagger]=\delta_{lj}$. Thanks to the linearity of the equation, we can separate the operator $\hat{s}$ into two contributions
\begin{eqnarray}
	\hat{s}_{l}(t)
	&= \sum_{m=1}^{N} J_{l m}(t)\hat{s}_{m}(0) + \sum_{\mu = \rightleftarrows}\int\nolimits_{0}^{\infty} F_{l,\omega,\mu}(t)\hat{a}_{\omega,\mu}(0), =: \hat{b}_{l}(t) + \hat{s}^\mathrm{noise}_{l}(t),\label{Eq:BosonAnsatz}
\end{eqnarray}
one that comes from the initial quantum and classical fluctuations in the waveguide, $\hat{s}^\mathrm{noise}(t)$, and which only acts on the waveguide modes, and another operator that describes all expected values from the cavities, $\hat{b}_{l}(t)$, and which coincides with the cavity operators at the beginning of the experiment $\hat{b}_{l}(0)=\hat{s}_l$.

The separation of operators from Eq.~\eqref{Eq:BosonAnsatz} will be central to this work and must be explained. As it is written, the first operator only acts in the emitters' subspace, while the second operator always contains an annihilation operator as right-most factor. From this structure it follows that, if the waveguide was initially empty, all expectation values that only involve the emitters can be written in terms of $\hat{b}_{l}$ operators. Specifically, using $[\hat{s}_m(t),\hat{s}_{l}^{\dagger}(t)]=\delta_{l,m}$, we can rewrite said expected values in normal order and use the fact that $\hat{a}_{\omega,\mu}(0)\ket{\psi(0)}=0$, to write, for instance
\begin{eqnarray}
	\braket{\hat{s}_{l}(t) \hat{s}_m(t)^\dagger}
	&= \braket{\hat{s}_{m}^{\dagger}(t)\hat{s}_{l}(t) } + \delta_{ml}=\braket{\hat{b}_{m}^{\dagger}(t)\hat{b}_{l}(t) } + \delta_{ml},\nonumber\\
	\braket{\hat{s}_{l}^{\dagger}(t) \hat{s}_{m}^{\dagger 2}(t)}
	&= \braket{\hat{s}_{m}^{2}(t) \hat{s}_l(t)}^* = \braket{\hat{b}_{m}^{2}(t) \hat{b}_l(t)}^*.\nonumber
\end{eqnarray}

This simplification is very useful, because the emitter operators can be expressed in a self-consistent way, without any reference to the waveguide operators. It also reflects the reality of many waveguide-QED experiments where the waveguide is initially empty. However, if this was not the case, we would simply need to solve the set of integro-differential equations that $F_{l,\omega,\mu}(t)$ obey as well.

The propagator satisfies $J_{\ell m}(t) = [\hat{b}_{\ell}(t),\hat{b}^{\dagger}_{m}(0)]$ and is the solution of the integro-differential equation
\begin{eqnarray}
	\frac{d}{dt} J_{\ell m}(t) &= \left[\frac{d}{dt}\hat{b}_{\ell}(t),\hat{b}_{m}(0)\right] =- \sum_{j=1}^{N} \int_{0}^{t} K_{lj}(t-\tau)e^{i\Delta(t-\tau)} J_{n m}(\tau) d \tau, \label{Eq:Jcoeff}
\end{eqnarray}
The problem's complexity reduces to solving $N$ coupled time-delayed differential equations~\eqref{Eq:Jcoeff}, which, as shown in Sect.~\ref{sec:memory-function}, can be simplified into a set of delay-differential equations that are easier to solve.

\subsection{Time-delayed interactions for two-level systems}
\label{sec:dde-tls}
Let us now discuss how the same theory applies to saturable emitters, by considering a system of $N$ identical two-level emitters (spin-like systems). In absence of environment, we could identify the emitter operator with a time-evolved Pauli ladder operator, $\hat{s}_{l}(t)=\hat{U}^{\dagger}(t)\hat{\sigma}^-_{l} \hat{U}(t)$ with $\hat{\sigma}_{l}^{-}(0)=(\ketbra{0}{1})_{l}$. In this case, the commutator $\hat{s}_{l}^0(t)=[\hat{s}_{l}^\dagger,\hat{s}_{l}]$ becomes the emitter polarization $\sim \hat{U}(t)^\dagger\hat{\sigma}^z_{l} \hat{U}(t)$ but can also be written as the nonlinear function $\hat{s}_{l}^0 = 2 \hat{s}_{l}^\dagger \hat{s}_{l} - \openone$  in terms of only $\hat{s}_{l}$. However, in the bosonic network we cannot write down an equation only for the $\hat{s}_{l}(t)$ operators, because of the nonlinear coupling $\hat{s}^0_{l}(t)\hat{\xi}_{l}(t)$ describing how the waveguide fluctuations at the beginning of the experiment affect the emitters.

Despite this apparent limitation, a Dyson series expansion of Eq.~\eqref{eq:integro_dif} shows that the lowest order contribution to $\hat{s}_{l}(t)$ is the sum of an operator that only contains qubit terms $\{\hat{\sigma}^-_{m}(t)\}_{m=1}^N$ at different times, plus a linear contribution from the waveguide that has a sum of annihilation operators $\hat\xi_{l}(t)$. This motivates us to extend the ansatz from Eq.~\eqref{Eq:BosonAnsatz} also to the saturable emitters, postulating a separation into purely emitter operators, $\hat{\sigma}^{-}_{l}(t)$, and a contribution from the waveguide's fluctuations
\begin{equation}
	\hat{s}_{l}(t) \simeq \hat{\sigma}^-_{l}(t) + \hat{s}_{l}^\mathrm{noise}(t). \label{eq:approximation_qubits}
\end{equation}
This separation is done with the assumption that, if the waveguide starts in a vacuum state, i.e. $\hat{a}_m(0)\ket{\psi(0)}=0\,\forall m$, any normal-order expectation value is approximated exclusively by a combination of those extended qubit operators:
\begin{eqnarray}
	&  \braket{\normord{F(\hat{s}_1(t),\ldots \hat{s}_N(t),\hat{s}_1^\dagger(t),\ldots,\hat{s}_N^\dagger(t))}} \label{eq:approx_expected_values} \\
	&\quad \quad= \braket{G(\hat{s}_1^\dagger(t),\ldots,\hat{s}_N^\dagger(t),\hat{s}_1(t),\ldots \hat{s}_N(t))}\nonumber\\
	&\quad \quad= \braket{G(\hat{\sigma}^+_1(t),\ldots,\hat{\sigma}^+_N(t),\hat{\sigma}^-_1(t),\ldots \hat{\sigma}^-_N(t))}.\nonumber
\end{eqnarray}

This decomposition implies, in particular, that the qubit matrix component is the operator that results from tracing out over the photonic vacuum degrees of freedom
\begin{equation}
	\hat{\sigma}^-_{l}(t) = \mathrm{tr}_\mathrm{vac}\hat{s}_{l}(t), 
\end{equation}
and that these operators satisfy an \textit{approximate} dynamical equation that no longer contains the initial state of the waveguide
\begin{equation}
	\frac{d\hat{\sigma}_{l}^{-}(t)}{dt} = \hat{\sigma}^{z}_{l}(t) \sum_{j=1}^{N} \int_{0}^{t} K_{lj}(t-\tau)e^{i\Delta(t-\tau)} \hat{\sigma}_{j}(\tau) d \tau.  \label{Eq:sigmaminusFinal}
\end{equation}
With the usual replacement $\hat{\sigma}^z = 2(\hat{\sigma}^-)^\dagger \hat{\sigma}^- - \openone$, this becomes a set of self-contained nonlinear equations, solvable over a space of complex matrices $\hat{\sigma}^-_{l}(t)\in\mathbb{C}^{2^N\times 2^N}$ that grows exponentially with the number of emitters.

Eq.~\eqref{Eq:sigmaminusFinal} generalizes previous works from literature describing single-photon dynamics in a system of $N$ emitters. Indeed, in the limit in which we have at most one excitation travelling in the system and the waveguide is initially in a vacuum state, this equation becomes exact with the replacement $\hat{\sigma}_{l}^-\to\braket{\hat{\sigma}_{l}^-} = c_{l}\in\mathbb{C}$. However, as we will show below, Eq.~\eqref{Eq:sigmaminusFinal} provides a very accurate description of the dynamics when compared with other exact methods, such as the multi-photon Wigner-Weisskopf theory.

It must be remarked that  Eqs.~\eqref{eq:approximation_qubits} and ~\eqref{eq:approx_expected_values} are the only essential approximation in this work. Unlike the bosonic model, these approximations are only justified when the waveguide is initially in a vacuum state---though it can be probably generalized to coherent states, with suitable displacements. This is per se not a terrible approximation as it is a good approximation of many useful setups and experiments

\subsection{Memory function kernel}
\label{sec:memory-function}
So far we have derived a family of integro-differential equations for the emitters that are approximate in the saturable case (qubits) and exact in the linear case (cavities). Still, the integral form makes the integration more difficult and hides important information, such as the explicit timescales for photon generation and interactions between quantum nodes. In ordinary quantum optical systems, further simplifications would transform these equations into a time-local Markovian model. Interestingly, while the bosonic networks described by~\eqref{eq:the-model} are obviously time non-local---i.e., information dropped into the network by one system may be recaptured by one or more quantum nodes at any future time---, the optical network is still amenable to the \textit{first} Born-Markov approximation. This approximation states that the dynamics of the environment in contact with the emitter is faster than the emitters' relaxation itself. In that scenario, the memory functions $K_{lj}(t)$ will become a combination of Dirac-delta functions, transforming the integro-differential equation into a delay-differential equation.

A common way to address the first Markov approximation is to state that the coupling between the emitters and the bosonic modes is approximately constant over the bandwidth of the photons that can be actually produced by the emitters---i.e., $|V_{n,\mu}(\omega)|^2$ changes very little or nothing over $[\Delta_l-\gamma_k,\Delta_l+\gamma_k]$ where $\gamma_l$ will be the bandwidth of those photons. As a first example, let us assume an infinite waveguide with $N$ emitters, imposing a uniform coupling that retains information about the emitters' positions
\begin{eqnarray}
	V_{n,\mu}(\omega) = g_{0} e^{i x_{n} \omega / v_{\mu}}. \label{Eq:NPoint-Like-V}
\end{eqnarray}
Introducing the group velocity $v_\mu$ and the time separations among emitters $\tau_{lj} = |x_{j} - x_{l}|/v_{\mu}$, and using $ \mathcal{PV} $ for Cauchy's Principal Value.
\begin{eqnarray}
	K_{lj} (u)
	&= \int\nolimits_{0}^{\infty} \left[|g_{0}|^{2} e^{-i\omega(u + \tau_{lj} )} +  |g_{0}|^{2} e^{-i\omega(u- \tau_{lj} )}\right] d\omega \nonumber\\
	&= K(u+\tau_{ij}) + K(u-\tau_{ij}) \label{eq:peaks}\\
	&=|g_{0}|^{2} \left[ \pi\delta(u+\tau_{ij}) - i \mathcal{PV}\left(\frac{1}{u+\tau_{ij}}\right) \label{eq:kernel_general}\nonumber \right. \\
	& \left. \quad\quad\quad\quad+\pi\delta(u-\tau_{ij})- i \mathcal{PV}\left(\frac{1}{u-\tau_{ij}}\right) \right].
\end{eqnarray}

\begin{figure}[t!]
	\centering
	\includegraphics[width=0.65\columnwidth]{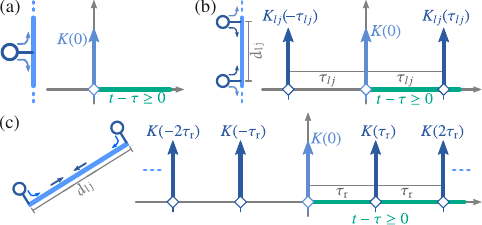}
	\caption{Illustration of the memory functions for different systems of emitters and waveguides. (a) Single emitter in an infinite waveguide. (b) two emitters in an infinite waveguide (c) two emitters at the ends  of a finite-length waveguide , }
	\label{Fig-MemoryFunction}
\end{figure}

The kernel function $K_{lj}(t)$ is the sum of two very narrow peaked functions~\eqref{eq:peaks} centered on $u=\tau_{ij}$ and $u=-\tau_{ij}$ (respectively $\tau = t-\tau_{lj}$ and $\tau=t+\tau_{lj}$). As sketched in Fig.~\SubFig{Fig-MemoryFunction}{b}, only one of those two peaks falls within the integration interval $\tau\in[0,t)$ in Eq.~\eqref{eq:integro_dif}. Thus, when emitter's operators $\hat{s}_l(t)$ and $\hat{s}_l(t)^\dagger$ evolve slowly over the relevant peak---i.e., \textit{locally Markovian dynamics}---we can approximate those operators as constants, simplifying the integro-differential terms
\begin{eqnarray}
	\label{eq:local-approx}
	&\int_0^t K_{lj}(t-\tau) e^{i\Delta_l t-\Delta_j\tau}\hat{s}_j(\tau)\mathrm{d}\tau \\
	&\quad\quad\simeq  z_{lj} e^{i(\Delta_j-\Delta_j)t}\hat{s}_j(t-\tau_{ij}),\mbox{ if } \tau_{ij}\in[0,t],\nonumber\\
	&\quad\quad = 0,\mbox{ else.}\nonumber
\end{eqnarray}
The real and imaginary parts of the complex number $z_{lj}$ describe dissipative and coherent processes~\cite{gonzalez-tudela2011, gonzalez-tudela2015, diaz-camacho2015}
\begin{equation}
	\label{eq:dynamical-parameters}
	z_{jj} = \frac{1}{2}\gamma_j - i \Delta^\mathrm{LS}_j,\mbox{ and } z_{l\neq j} = \frac{1}{2}\gamma_{l j} -i g_{l j},
\end{equation}
with the emitters' spontaneous decay rate, $\gamma_j$, the Lamb-shift induced by the interaction with the environment,  $\Delta^\mathrm{LS}_j$, the collective dissipation $\gamma_{l\neq j}$ the coherent photon exchange between emitters $g_{l\neq j}$.

The first Markov approximation can be used to transform the integro-differential equation Eq.~\eqref{Eq:sigmaminusFinal}, into a set of delay-differential equations for operators ($\Delta_j=\Delta$). This is done by transforming the equations into functional form, with the help of Heaviside functions  $\Theta(t-\tau_{lj})$ that only activate when $\tau_{lj}\in[0,t]$
\begin{equation}
	\frac{d\hat{s}_{l}(t)}{dt} = \hat{s}^{0}_{l}\hat{\xi}_{l}(t) + \hat{s}^{0}_{l} \sum_{j=1}^{N} z_{l j} \hat{s}_{j}(t-\tau_{lj})\Theta(t-\tau_{lj}).
\end{equation}
Combined with the exact~\eqref{Eq:Jcoeff} or approximate~\eqref{Eq:sigmaminusFinal} elimination of the waveguide $\hat{\xi}$, this leads to a set of standalone delayed-differential equations for the operators for the emitters, which can be integrated numerically (c.f. Sect.~\ref{sec:validation}).

This reasoning can be generalized to finite links and complex networks (c.f. Figs.~\SubFig{fig:setup}{a} and Fig.~\SubFig{fig:setup}{c}), but now the set of delays developes a more sophisticated structure
\begin{equation}
	K_{lj}(u) = \sum_{n=0}^{\infty} \sum_{r}K(u- n \tau_{lj}^{(r)}).
\end{equation}
Pairs of emitters can exchange photons via multiple paths, depending on the network's topology, and each path involves a possibly different delay $\tau_{lj}^{(r)}$.

Finally, let us emphasize that the first Markov approximation is useful to convert the dynamical equations into a DDE form. However, this is not strictly required and one could consider the case of more general memory functions, paying the price of introducing additional dynamical variables to keep track of the photonic currents.

\section{Applications and exact limits}
\label{sec:applications}
So far, Sects.~\ref{sec:dde-cavity}-\ref{sec:dde-tls} developed two formal sets of HL equations, with prescriptions to transform the time non-locality into a discrete delay-differential form (Sect.~\ref{sec:memory-function}). The following sections will apply this theory to different problems and models. The goal is two-fold: to prove that the methodology is consistent with the literature of Markovian and non-Markovian models and to develop explicit equations used in benchmarks (c.f. Sect.~\ref{sec:validation}) and in the exploration of new physical predictions associated to time-delayed collective phenomena (c.f. Sect.~\ref{sec:cascaded}).

\subsection{Exact limit: single excitations}
\label{sec:single-excitations}
The model in Eq.~\eqref{eq:the-model} conserves the number of excitations in the qubits and the waveguide, that is $[\hat{N},\hat{H}]=0$ for
\begin{equation}
	\hat{N} = \sum_j \hat{s}_j^\dagger\hat{s}_j + \sum_k \hat{a}_k^\dagger \hat{a}_k.
\end{equation}
If we start with an empty waveguide and at most $N$ excited emitters and we restrict ourselves to a finite set of collective modes in the optical network, it is possible to simulate the dynamics of the whole system (waveguide plus emitters) using a Wigner-Weisskopf ansatz with exactly $\hat{N}$ excited objects. For $N=1,2$ these read
\begin{eqnarray}
	\ket{\psi_1(t)}
	& = \left[\sum_j c_j(t) \hat{\sigma}^+_j  + \sum_k \psi_k(t) \hat{a}_k^\dagger\right]\ket{\mathrm{vac}},\label{eq:WW-1excitation}\\
	\ket{\psi_2(t)}
	&=\left[\sum_{l j} c_{l j}(t) \hat{\sigma}^+_j\hat{\sigma}^+_l  + \sum_{kr} \psi_{kr}(t) \hat{a}_k^\dagger \hat{a}_r^\dagger\right]\ket{\mathrm{vac}} \nonumber \\
	&+\sum_{jk}\chi_{jk}\hat{\sigma}^{+}_{j}\hat{a}_k^\dagger\ket{\mathrm{vac}}. \label{eq:WW-2excitation}
\end{eqnarray}
In these exact representations, the Schrödinger equation becomes a set of sparsely coupled ODEs for the vectors of coefficients $\vec{v}_1 \in\mathbb{C}^{M+K}$ and $\vec{v}_2\in\mathbb{C}^{M(M-1)/2+K(K-1)/2+KM}$, for $M$ emitters and $K$ waveguide modes. These equations may be accurately integrated using approximate exponentiation techniques, to compute any emitter and photon properties~\cite{penas2022, penas2024}, which is the method used in Sects.~\ref{sec:quantum-link-benchmark} and~\ref{Sect:2LevelInf} to benchmark our model.

The case of single-photon dynamics is particularly interesting. In that limit, we can derive ODEs for the emitter coefficients that mimic those derived for the operators $\hat{\sigma}^-_{l}$. In the limit $\psi_k(0)=0$, these equations read
\begin{equation}
	\frac{dc_{l}^{-}(t)}{dt} = -  \sum_{j=1}^{N} \int_{0}^{t} K_{lj}(t-\tau) e^{i(\Delta_{l}t-\Delta_{j}\tau)}c_{j}(\tau).
	\label{eq:integro_diff-WW}
\end{equation}
This equation is identical to the linear model~\eqref{Eq:Bcavity}, because quantum oscillators and qubits are indistinguishable in a subspace with at most one excitation. More interestingly, this equation is also exactly the outcome of projecting Eq.~\eqref{eq:integro_dif} onto a quantum state with one excitation~\eqref{eq:WW-1excitation}, verifying that, at least in this limit, \textit{our methodology is exact}. Finally, Eq.~\eqref{eq:integro_diff-WW}, subject to the same considerations as in Sect.~\ref{sec:memory-function}, becomes a delay-differential equation for the $c_{l}(t)$ coefficients, similar to earlier models for single emitters in a semi-infinite waveguide~\cite{dorner2002, tufarelli2013, tufarelli2014} and other non-Markovian problems.

\subsection{Multiple emitters in an infinite waveguide}
\label{sec:infinite-wavegudie-theory}
As a first application, let us consider a system of $N$ two-level identical emitters, with natural frequencies $\Delta$, coupled to an infinite waveguide, as depicted in Fig.~\ref{fig:setup}b. Using the tools from Sect.~\ref{sec:theory}, and in particular the combination of the approximate model Eq.~\eqref{Eq:sigmaminusFinal} with the explicit form for the kernel function, we recover a set of \textit{time-delayed integro-differential} HL equations
\begin{eqnarray}
	\frac{d}{dt} \hat{\sigma}^{-}_{l}(t)
	&= \hat{\sigma}^{z}_{l}(t) \sum_{n=1}^{N} G_{nl}
	\hat{\sigma}_{n}^{-}\left( t - \tau_{nl}\right) \Theta\left( t - \tau_{nl}\right) , \label{Eq:sigmaInfiniteWG} 
\end{eqnarray}
with a Green function tensor that is consistent with the 1D-waveguide models~\cite{gonzalez-tudela2011, shi2015}
\begin{eqnarray}
	G_{nl} =  \frac{\sqrt{\gamma_n\gamma_{l}}}{2} e^{i k_{0}  |x_{n} -x_{l}| } + i \Delta^\mathrm{LS}_n\delta_{nl}, \label{Eq:GreenTensor}
\end{eqnarray}
where $k_0$ is the quasimomentum associated to the photons of frequency $\sim\Delta$ created by the emitters. As before, $\gamma_{n}$ and  $\Delta^{\mathrm{LS}}_n$ represent the spontaneous emission rate and the Lamb shift of the individual emitters, respectively, which are also related to the coherent and incoherent interactions, respectively. Finally, using the identity $\hat{\sigma}^z_{l}(t) = 2(\hat{\sigma}^-_{l}(t))^\dagger \hat{\sigma}^-_{l}(t)-\openone$, we can transform this model into a set of standalone equations for $N$ complex matrices $\hat{\sigma}^-_{l}(t)\in\mathbb{C}^{2^N\times 2^N}$.

\subsection{Exact limit: Markovian master equation}
\label{sec:master-equation}
In some studies we wish to compare our time-delayed HL equations with the physics of emitters in the limit of zero delays, in which $\gamma_{\ell}\tau_{\ell j}$ and $\gamma_{j}\tau_{\ell j}$ can be neglected. In that limit, the emitters can be modeled using a Markovian master  equation~\cite{gonzalez-tudela2015}
\begin{eqnarray}
	\frac{d \hat{\rho}(t)}{dt}
	&=  \frac{1}{2}\sum_{l,j=1}^{N} \Gamma_{jl} \left[ 2 \hat{\sigma}_{l}^{-} \hat{\rho}(t) \hat{\sigma}_{j}^{+} - \{\hat{\sigma}_{j}^{+}\hat{\sigma}_{l}^{-},\hat{\rho}(t) \} \right]
	\nonumber \\
	&+ i \sum_{l,j=1}^{N} [\Delta_{jl}^\mathrm{LS} \hat{\sigma}^{+}_{l} \hat{\sigma}^{-}_{j}, \hat{\rho}(t)]. \label{Eq:MasterEquation}
\end{eqnarray}
The matrices $\Gamma$ and $\Delta$ describe the incoherent and coherent interactions between emitters, $\Delta_{jl}^\mathrm{LS}=\sqrt{\gamma_{j}\gamma_{l}} \sin(k_{0}  |x_{j} -x_{l}|)/2$ and $\Gamma_{jl} = \sqrt{\gamma_{j}\gamma_{l}} \cos(k_{0}  |x_{j} -x_{l}|)$, as well as the Lamb-shifts and spontaneous emission rates $\Gamma_{jj}=\gamma_j$ and $\Delta_{jj}^\mathrm{LS}=\Delta_j^\mathrm{LS}$. These matrices depend on the topology and separation between the emitters along different paths.

For resonant two-level systems in a 1D infinite waveguide~\cite{gonzalez-tudela2011, shi2015}, in particular, we have
\begin{eqnarray}
	G_{l j} = \frac{1}{2}\Gamma_{l j} + i\Delta_{l j},
\end{eqnarray}
consistently with the non-Markovian couplings~\eqref{Eq:GreenTensor}. Furthermore,as shown in appendix ~\ref{Sec:appendix_b}, when computing the equations for expected values
\begin{equation}
	\braket{\hat{\sigma}^-_l} = \mathrm{tr}\{\hat{\sigma}^-_l\hat{\rho}(t)\} ,
\end{equation}
one arrives at the same HL model from Eq.~\eqref{Eq:sigmaInfiniteWG}, with the substitution $\tau_{jl}=0$.
We thus conclude that the master equation for multiple emitters in a waveguide and the zero-delay limit of our HL theory are equivalent. However, in this limit, we must recognize that it is more convenient to solve a single equation for $\rho(t)\in\mathbb{C}^{2^N\times 2^N}$ using Eq.~\eqref{Eq:MasterEquation}, rather than $N$ problems of comparable size for all the matrices $\{\hat{\sigma}^-_{l}(t)\}_{l=1}^N$.

\subsection{Quantum link with two nodes}
\label{sec:quantum-link-theory}

As a second and very relevant application, let us consider a finite-length waveguide interfacing two quantum nodes, each connected at a different end of the waveguide, as illustrated in Fig.~\SubFig{fig:setup}{a} and Fig.~\SubFig{Fig-MemoryFunction}{c}. This model is the simple instantiation of a quantum link, a setup studied experimentally~\cite{zhong2018, leung2019, magnard2020, chang2020, zhong2021, storz2023, niu2023, grebel2024, qiu2025} and theoretically~\cite{xiang2017, penas2022} as a model for a quantum processor interconnect.

As discussed in Sect.~\ref{sec:memory-function} and depicted in Fig.~\SubFig{Fig-MemoryFunction}{c}, a single photon created by one emitter can travel multiple times along the waveguide, regardless of whether it is reabsorbed or not. The memory function thus becomes an inifinite train of narrow peaks $K(u)$ with spacings commensurate with the return time $\tau_{12}=d_{12}/v$ of a photon travelling for a distance $d_{12}$
\begin{eqnarray}
	K_{jj}(u) & \sim z_{jj}\sum_{n=0}^{\infty} \delta (u-2n \tau_{12} ),\label{Eq:close_Memory}\\
	K_{12}(u) & \sim z_{12}\sum_{n=0}^{\infty} \delta (u -2n \tau_{12} ).\nonumber
\end{eqnarray}
Note that, out of the infinite train, only peaks that fall within the integration interval $[0,t]$ contribute to the dynamics. This gives a total of $\lfloor{t / \tau_{12}}\rfloor$ non-local contributions with complex weights that repeat cyclically in time. Still, as shown in Sect.~\ref{sec:memory-function}, this once more transforms Eq.~\eqref{eq:integro_dif} into a set of amenable, delay-differential equations that can be integrated.

Replacing the memory function~\eqref{Eq:close_Memory} into Eq.~\eqref{eq:integro_dif} and assuming identical two-level systems at the ends of the quantum link produces
\begin{eqnarray}
	\frac{d\hat{\sigma}^{-}_{1}(t)}{dt}  &= -\left(i \Delta^\mathrm{LS} +\frac{\gamma}{2}\right)\hat{\sigma}^{-}_{1}(t) + i\sqrt{\gamma}\hat{\sigma}^z_1(t)\hat{\xi}^\mathrm{in}_1(t) \\
	\frac{d\hat{\sigma}^{-}_{2}(t)}{dt}  &= -\left(i \Delta^\mathrm{LS} +\frac{\gamma}{2}\right)\hat{\sigma}^{-}_{2}(t) + i\sqrt{\gamma}\hat{\sigma}^z_2(t)\hat{\xi}^\mathrm{in}_2(t).\nonumber
\end{eqnarray}
The first term in each equation describes the self-evolution of the emitter in the waveguide, much as what one would expect from a Markovian theory. The second term, on the other hand, illustrates the reaction of the qubit to a stream of photons
\begin{eqnarray}
	& \hat{\xi}^\mathrm{in}_{\ell}(t) =
	-\sum_{m_e=2,4,\ldots} \sqrt{\gamma} \hat{\sigma}^{-}_{\ell}(t - m_e \tau_{12} ) \Theta(t - m_e\tau_{12}) \nonumber\\
	&~~-  \sum_{m_o=1,3,\ldots} \sqrt{\gamma}\hat{\sigma}^{-}_{3-\ell}(t - m_o\tau_{12}) \Theta(t - m_o\tau_{12}) e^{ik_0L}
\end{eqnarray}
that includes both photons emitted by the opposite node, as well as the photons created in this node and reflected multiple times in the past.

Thanks to both emitters being identical, they both acquire similar Lamb shifts $\Delta^\mathrm{LS}$ and decay with the same spontaneous emission rate $\gamma$. Furthermore, this emission rate $\gamma$ dictates both the strength of the photon stream that arrives at each node, $\hat{\xi}^\mathrm{in}$, as well as the strength of the coupling of the emitters to that stream, $i\sqrt{\gamma}\hat{\sigma}^z_1(t)\hat{\xi}^\mathrm{in}_1(t)$. Finally, the phase factor $e^{ik_0L}$ reflects the optical path of photons travelling along the waveguide and is $e^{ik_0L}\simeq \pm 1$ for emitters resonant to any of link's modes.

\section{Collective dynamics and benchmarks with two photons}
\label{sec:validation}
So far, we have found a non-local HL theory that is exact in the linear limit (Sect.~\ref{sec:dde-cavity}), in the subspace of one excitation (Sect.~\ref{sec:single-excitations}) and in the Markovian limit of negligible delays (Sect.~\ref{sec:master-equation}). However, we also wish to explore the application of these equations to regimes where ~Eq.\eqref{eq:approximation_qubits} should be regarded as a heuristic approximation whose accuracy is not guaranteed --- that is, systems with many photons and non-negligible retardations.

To enable this exploration, we have performed benchmarks of the DDE model against exact diagonalizations of problems with few excitations, which can be treated exactly using the Wigner-Weisskopf ansatz from Eqs.~\eqref{eq:WW-1excitation}-\eqref{eq:WW-2excitation}. The following sections describe both the numerical methods we use to integrate the HL equations in the emitters' subspace, as well as specific results with multiple emitters in infinite and finite waveguides.

\subsection{Numerical integration of the DDE Heisenberg-Langevin model}
\label{eq:numerical-method}
The dynamics of the emitters in the time non-local models from Eqs.~\eqref{Eq:sigmaInfiniteWG} or \eqref{Eq:close_Memory} presents two unavoidable complications. First, in the worst case of saturable emitters, the unknowns of our problem, $\hat{s}_{l}\sim \hat{\sigma}^-_{l}$, are complex matrices whose size grows exponentially with the number of emitters. At this stage, we will not discuss the size limitation imposed by the size of the operators $\hat{\sigma}^-_{l}$, which can be addressed using tools such as tensor network representations and matrix product operators.

The second, but arguably simpler complication, is present in both the linear~\eqref{Eq:Jcoeff} and the nonlinear models~\eqref{Eq:sigmaminusFinal}. This is the fact that both equations have a non-local structure
\begin{equation}
	\frac{d}{dt}X = f(X(t),t) + g(X(t),t) \sum_n h_n(X(t-\tau_n))\Theta(t-\tau_n),
	\label{eq:non-local-dde}
\end{equation}
with a source term that depends on delayed values of the operators or matrices in the past, $X(t-\tau_n)$. We address this challenge by working in a rotating frame in which we have eliminated the qubit's intrinsic frequencies. In that frame, the unknowns $X\sim J(t)$ or $\hat{\sigma}^-_{l}$ are smooth functions, that can be sampled at regular intervals of time. Then, using linear or spline interpolation, we can reconstruct the current operators that appear in Eq.~\eqref{eq:non-local-dde}, to compute the instantaneous time-derivative.

This technique may be combined with different types of explicit Runge-Kutta-like solvers to achieve decent accuracy. The result is a numerical method that scales as $O(d^2)$ with respect to the dimensionality of our quantum system---i.e., $X\in\mathbb{C}^{d\times d}$ with $d=2^N$ for $N$ qubits, or $X\in\mathbb{C}^{N\times N}$ and $d=N$ for bosons---. This  performance is very competitive with ealier methods based on tensor networks, with scalings ranging from $\mathcal{O}(d^3)$ to $\mathcal{O}(d^{12})$ for time evolution, as compared in Ref.~\cite{vodenkova2024}. Furthermore, in many situations with large numbers of emitters, we have found that the $X$ matrices have a very sparse structure, that lowers the cost closer to $\mathcal{O}(d^1)$, with a prefactor that depends on the time step and interpolation method $\mathcal{O}(\Delta{t}^{-1})$. Finally, it must be emphasized that the complete method can be written in around 100 lines of standard Python code, without dependencies on tensor contraction methods, in a very accessible way that can be understood and reproduced by undergraduate students. An example of the code employed can be found in \cite{DDE_Code}

In the following sections, we will display results from those simulations, compared with numerically exact simulations of two-photon Wigner-Weisskopf equations~\eqref{eq:WW-2excitation}, both to test the accuracy of the method and to bootstrap the exploration of collective phenomena in arrays of quantum emitters.

\subsection{Analytic solution of bosonic emitters in an infinite waveguide}
\label{sec:bosons-analytic}
The bosonic problem can be used as a benchmark of the DDE solver strategy, because it admits a simple analytical solution. To illustrate this, we focus on the problem of $N$ emitters, placed with a uniform separation $d$ in an infinite waveguide. This setup is central to the physical discussions below, when we compare linear emitters against two-level systems in the context of super- and subradiant experiments (c.f. Sect.~\ref{sec:cascaded}).

\begin{figure}[t]
	\centering
	\includegraphics[width=0.65\linewidth]{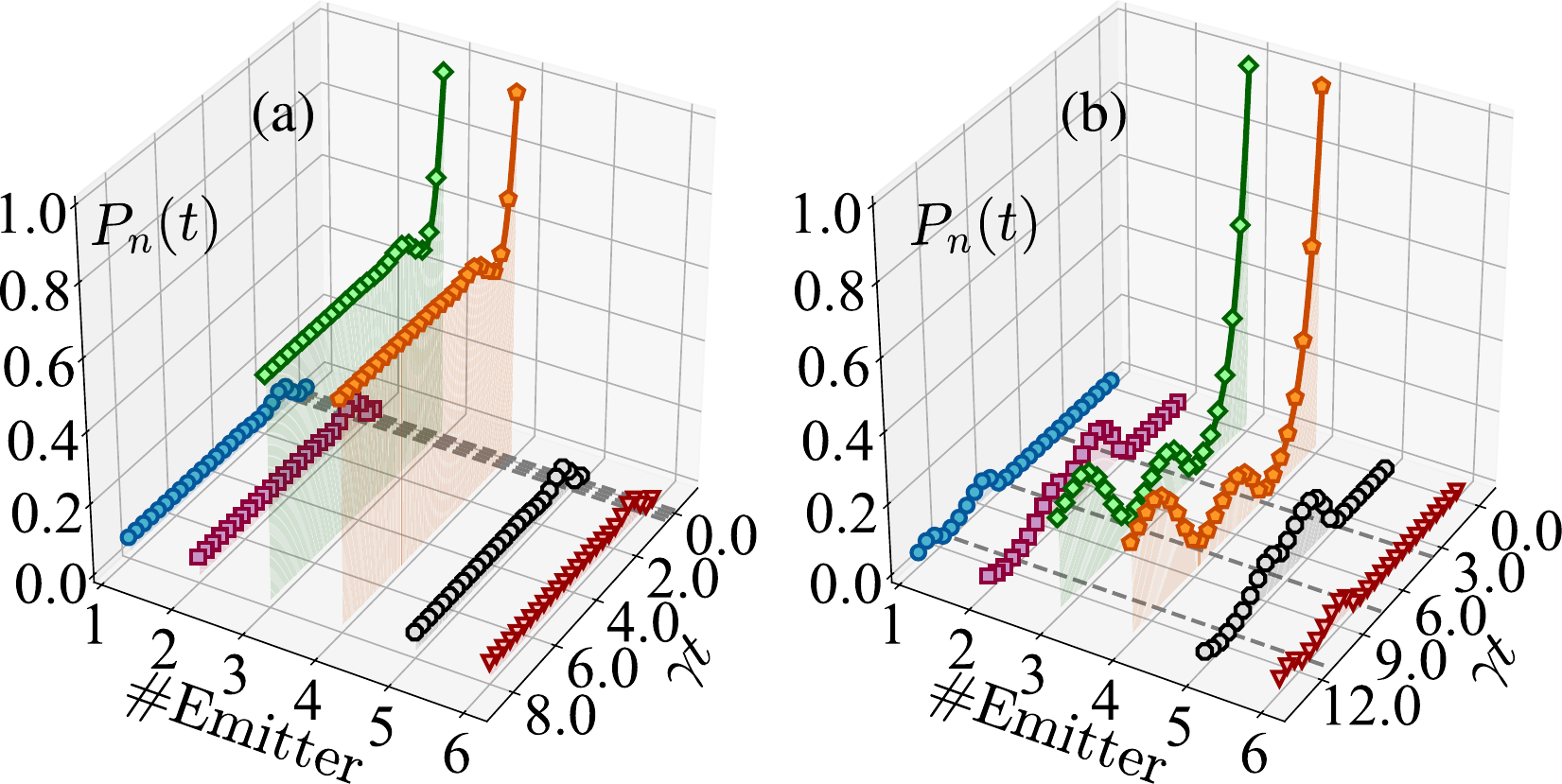}
	\caption{Population of the bosonic emitters as function of the dimensionless time $\gamma t$ and the emitters' label. For both simulations we have fixed $\phi_{0} = 2\pi$ and  $\ket{\psi(0)} = \ket{001100} $. In the subfigure (a) we compare a numerical integration (dot symbols) with the analytical solution (solid line), considering the nearly-Markovian regime $\gamma \tau_{12} = \pi / 20$. In the subfigure (b) we repeat the same comparison, now considering the Non-Markovian regime with $\gamma \tau_{12} = 5\pi/4 $, as we set $\phi_0=50\pi$.}
	\label{fig:wigner-weisskopf-cavity}
\end{figure}

In this particular scenario, the combination of the bosonic equations~\eqref{Eq:Jcoeff} with the previous analysis of the memory function~\eqref{eq:peaks} leads to a delay-differential equation for the matrix of coefficients $J(t)$, given by
\begin{eqnarray}
	\frac{d}{dt} J_{l m}(t) &=- \sum_{n=1}^{N} G_{nl} J_{n m}\left( t - \tau_{n\ell}\right) \Theta\left( t - \tau_{nl}\right), \label{Eq:Jcoeff_infinity}
\end{eqnarray}
with a matrix of coherent and incoherent interactions given by Eq.~\eqref{Eq:GreenTensor}, in the limit of identical tensors $\gamma_n=\gamma$ and negligible Lamb shift $\Delta^\mathrm{LS}=0$. In this particular scenario, the only free parameters are the optical phase acquired by photons travelling among neighboring emitters, $\phi_0 := k_0|x_{n+1}-x_{n}|=k_0d$, and the adimensionalized decay fraction as photons travel among quantum nodes $\gamma \tau_{12} = \gamma d/v_g$.

As explained above, this equation can be solved numerically by sampling the matrix $J(t)$ at regular intervals and interpolating those values to estimate the non-local current in Eq.~\eqref{Eq:Jcoeff_infinity}. In this particular instance, however, it is also possible to compute the explicit solution to $J_{\ell m}(t)$ using a Laplace transform and geometric series, similar to earlier works in the context of waveguide-QED~\cite{sinha2020}. The analytical solution for $N$ emitters can be achieved by an algorithm that divides the total integration in time intervals of the form $[m\tau_{12},(m+1)\tau_{12}]$ and uses the solutions of the $J_{l n}(t)$ functions in previous intervals, starting with $J_{l n}(t) = \delta_{l n}$ when $t \in [0,\tau_{12}]$ (see~\ref{Sec:App:Solution} for further details).

Having an analytical solution to the bosonic problem allows us to gauge the accuracy of the interpolation strategy when solving the DDE equation. As an illustration of this, Fig.~\ref{fig:wigner-weisskopf-cavity} compares the dynamics of six bosonic emitters in an infinite waveguide, obtained using both methods. In the two simulations presented, there are six emitters of which two are initially in an excited state $\ket{\Psi(0)}=\hat{s}_3^\dagger(0) \hat{s}_4^\dagger(0)\ket{\mathrm{vac}}$, with no photons in the waveguide. The free spontaneous emission rate $\gamma$ for all emitters is the same, as well as the phase acquired by photons travelling between emitters, $\phi_0 = k_0d=2\pi$, but we compare a near Markovian $\gamma\tau_{12}=\pi/20$  (c.f. Fig.~\ref{fig:wigner-weisskopf-cavity}b) against a highly non-Markovian scenario, $\gamma\tau_{12}=5\pi/4$ (c.f. Fig.~\ref{fig:wigner-weisskopf-cavity}b), where the delays are very relevant and the accuracy of the integrator is very relevant.

As figures of merit for the comparison we use the population of the different bosonic modes. Since the waveguide is assumed to be in a vacuum state, and since the number of photons $\hat{n}_{l}(t)=\hat{s}_{l}^\dagger(t)\hat{s}_{l}(t)$ is already in normal order, this quantity can be simplified to an expected value over the emitter's operators
\begin{eqnarray}
	P_l(t)
	&:=  \braket{\Psi(0)|\hat{n}_{l}(t)|\Psi(0)}\\
	&=  \sum_{m=1}^{N}\sum_{n=1}^{N} J_{l m}^{\ast}(t)J_{l n}(t)\braket{\Psi(0)| \hat{s}^{\dagger}_{m}(0) \hat{s}_{n}(0) |\Psi(0)}.\nonumber
\end{eqnarray}
As shown in Fig.~\ref{fig:wigner-weisskopf-cavity}, the predictions of the numerical simulation agree with the analytical solution. These plots also evidence some of the physics that will be discussed later on, including the existence of a light cone that determines the excitation of emitters, the emergence of bound states for this particular phase relation, evident at $t > 5\tau_{12}$, and the reduction of the weight of these bound states as emitters become more and more separated.

\subsection{Two-level emitters in infinite waveguide}
\label{Sect:2LevelInf}
Our next benchmark focuses on the saturable emitters and the accuracy of the approximation from Sect.~\ref{sec:dde-tls}. As in the previous subsection, we assume a chain of $N$ regularly spaced emitters, with separation $d$ among nearest-neighbor and placed in an infinite waveguide. In the HL model with a discrete memory function~\eqref{Eq:sigmaInfiniteWG}, the relevant parameters are the adimensionalized spontaneous emission rates of individual qubits $\gamma\tau_{12}$ and the phase acquired by photons when they travel among emitters $\phi_0=k_0d$. This is done mainly to simplify the equations, since a more general case (different spacing) would request essentially the same amount of computational resources

\begin{figure}[t]
	\centering
	\includegraphics[width=0.65\linewidth]{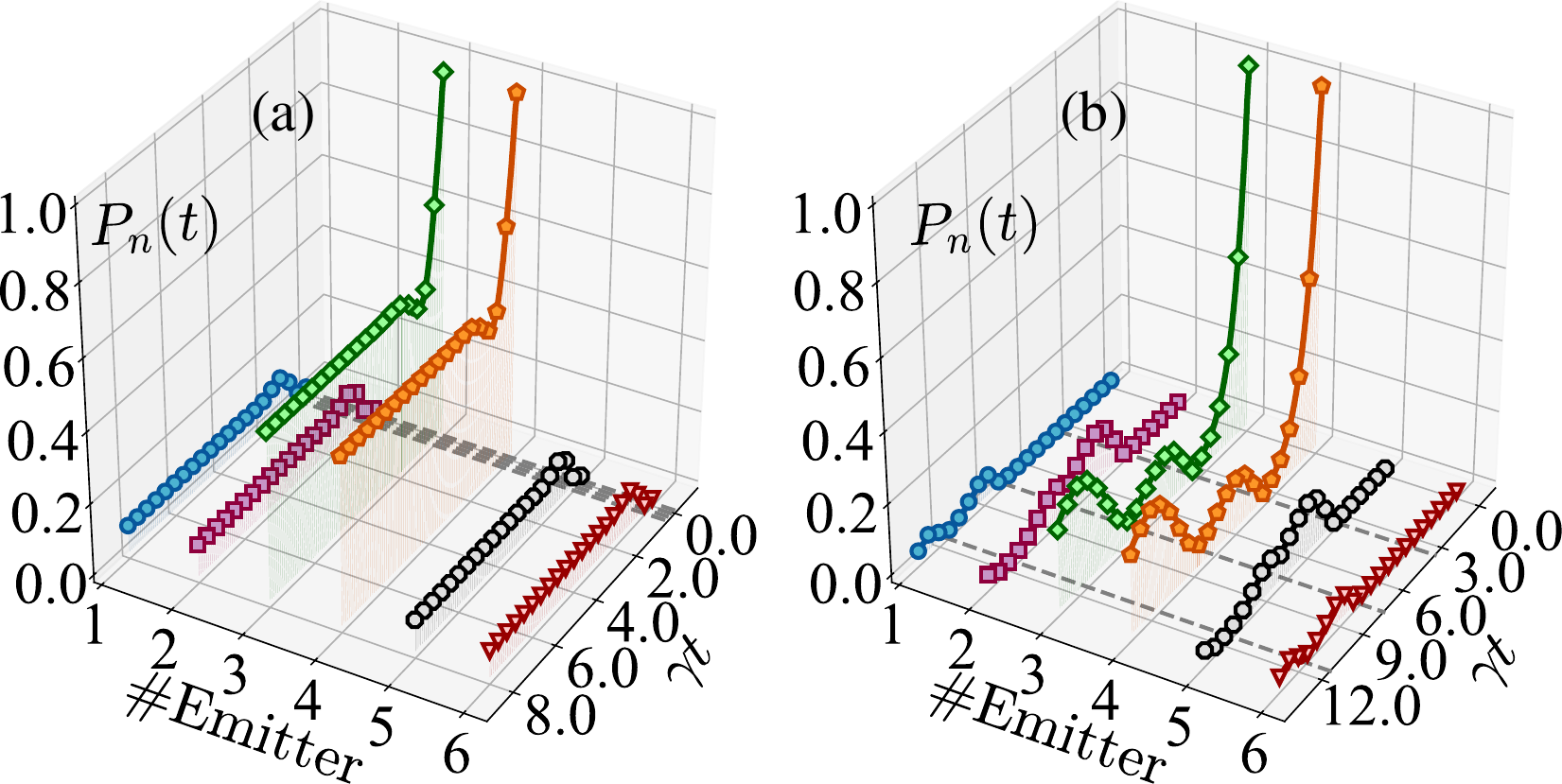}
	\caption{Population of the two-level  emitters' excited state $P_{n}(t)$ as a function of the dimensionless time $\gamma t$ and the emitters' label. Both simulations compare the outcome of the HLS equations (dot symbols) with an exact diagonalization of the Wigner-Weisskopf ansatz (solid line), either in (a) a nearly-Markovian regime $\phi_0=2\pi$, $\gamma \tau_{12}=\pi/20$ or in (b) a deeply retarded configuration $\phi_0=50\pi$, $\gamma\tau_{12}=5\pi/4$.}
	\label{fig:wigner-weisskopf}
\end{figure}

Unlike the bosonic case, there is not an exact solution which we can compare the HL model against. Instead, we have to trace back to the original Hamiltonian~\eqref{eq:the-model} of which the locally Markovian equations~\eqref{Eq:sigmaInfiniteWG} are an approximation. We can compare the HL model against a numerically exact simulation of the Wigner-Weisskopf ansatz with up to two excitations---i.e. Eqs.~\eqref{eq:WW-1excitation}-\eqref{eq:WW-2excitation}---as it evolves under a specific form for the Hamiltonian~\eqref{eq:the-model}.

In particular, the HL equations will be benchmarked against such simulations of the Wigner-Weisskopf model for a finite-length waveguide, considering a finite number of photon modes. That is, we do a matrix representation of the Hamiltonian ~\eqref{eq:the-model} in the joint Hilbert space of the emitters and the waveguide, but with a truncation in the modes of the waveguide. This allows us to simulate the system using Schrödinger's equation. To ensure good accuracy, the length of the waveguide is chosen to exceed the distance travelled by photons during the numerical experiments $L\gg v_g T$, with $T$ the total evolution time. To ensure a good, locally Markovian behavior of the emitters, the model assumes a linear dispersion relation $\omega_k = v_g k$, with equispaced modes $k=\pi/L\times \mathbb{N}$, describing the stationary modes. For the convergence of the simulations, we simulated photons over a bandwidth $20/\gamma$ using a long waveguide, $L/v_g > 40/\gamma$, with 500 and up to 1000 modes. The parameters for the HL model are fitted to the Wigner-Weisskopf simulation in the following way. First, the spontaneous emission of a single emitter in an empty waveguide is used to calibrate the rate $\gamma$. Then, for experiments with multiple emitters, the formula in Eq.~\eqref{Eq:GreenTensor} is used to determine the interaction between emitters.

With this information, both models are compared in simulations between $N=1$ to $N=8$ emitters, and up to two excitations, with similar results. As an example of the outcome of this study, Fig.~\ref{fig:wigner-weisskopf} shows two simulations with $N=6$ emitters, in which the initial state has one excitation in each of the two central emitters and the waveguide is in the vacuum state,  $\ket{\Psi(0)}=\ket{001100}\ket{\mathrm{vac}}$. As Sect.~\ref{sec:bosons-analytic}, simulations are parameterized by two dimensionless quantities, $\phi_0$ and $\gamma\tau_{12}$, which represent the phase acquired by a photon and the decay experienced by an emitter during one retardation period. We select the phase $\phi_{0}$ as a multiple of $2\pi$ that supports the emergence of photon bound states, and illustrate results with both a small delay, close to a Markovian regime $\gamma\tau_{12}=\pi/20$ (c.f. Fig.~\ref{fig:wigner-weisskopf}a), by setting $\phi_0 = 2\pi$, and with a significant retardation $\gamma \tau_{12} = 5\pi/4 $ (c.f. Fig.~\ref{fig:wigner-weisskopf}b), with $\phi_0 = 50\pi$.

As before, we compare the simulation methods by inspecting the emitters' excitation probability $P_l(t)=\braket{\hat{s}_l^\dagger(t)\hat{s}_l(t)}$. Fig.~\ref{fig:wigner-weisskopf} shows that the HL simulations (solid line) are in good agreement with the exact diagonalization model (dots).

Both the exact diagonalization and the HL equations accurately reproduce the emergence of a light cone, evident in the way that qubits influence each other by the exchange of photons. To make the interpretation of the time of arrivals more clear, both figures show the multiples of the retardation times $t=n\times \tau_{12}$ as grey lines. At $t=\tau_{12}$ we see in both figures how qubits 2 and 5 get excited due to the influece of the arriving photons, while at $t=2\tau_{12}$  qubits 1 and 6 begin a similar dynamics.

There are also two different physical regimes. In the near Markovian limit with short retardations, a bound state in the continuum is quickly formed, with a fraction of a photon mostly trapped between qubits 3 and 4, with some population in the remaining qubits. In the highly retarded non-Markovian case, in which separations are large, $\phi_0\sim 50\pi$ and $\gamma\tau_{12}\sim 5\pi/4$, the two central qubits decay very rapidly and a smaller fraction of a photon is trapped, leading to a non-stationary breathing between all qubits---which can probably be explained by a combination of BIC states.

\begin{figure*}[t!]
	\centering
	\includegraphics[width=1.0\linewidth]{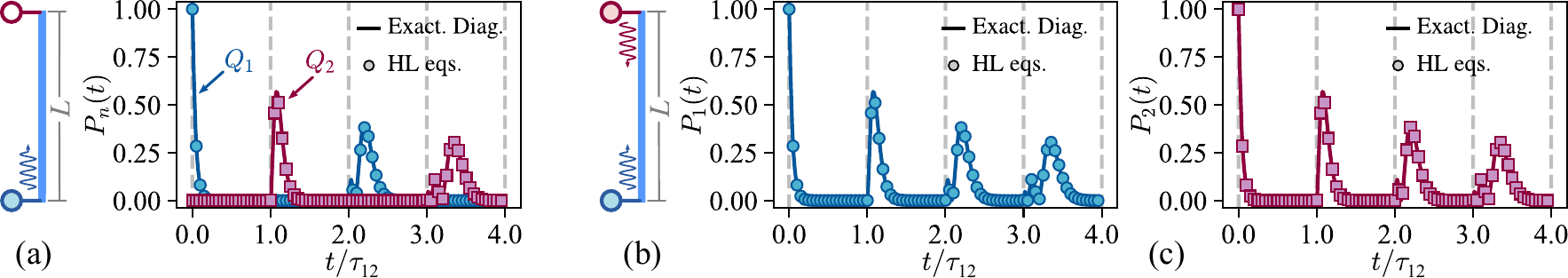}
	\caption{Population of the excited state $P(t)$ of two two-level emitters in a closed waveguide as a function of the adimensionalized time. The solid and dotted lines represent the results from the exact diagonalization and the HL equation methods, respectively. Colors blue and purple are used to denote observations on the first and second qubit, respectively. The dashed gray lines indicates the time interval $\tau_{12}$ needed for a photon to traverse the waveguide. In both simulations we have fixed $\phi_{0} = 2\pi $ and $\gamma \tau_{0} = 8 \pi  $. Fig.(a) shows the qubit's excitation probabilities as a function of time, for the initial state with one excited qubit $\ket{10}$. Figs. (b-c) show the same quantities, for an initial state with two excitations $\ket{11}$.}
	\label{fig:two_qubit_closewavguide}
\end{figure*}

Interestingly, in the near-Markovian regime Fig.~\ref{fig:wigner-weisskopf}a, already at $t=2\tau_{12}$ the bound state has been fully formed, which seems to contradict the intuition that, for correlations to establish and the system to be fully stationary, at least a time $t=5\tau_{12}$ is required---i.e. the time to establish a link between qubits 1 and 6---or $t=4\tau_{12}$---i.e., the time for emitters 3 and 4 to have ``probed'' the boundaries of the array. Furthermore, in the non-stationary regime, we can infer that $\tau_{12}$ is the relevant timescale for the breathing modes that are observed, suggesting that this physics could be due to a multiplicity of bound states with broken degeneracies of order $1/\tau_{12}$.

\label{discussion:errors}%
It must be remarked that the comparison between both models is not exact. The differences in the predictions by both models can be explained in three ways. First, there are errors that arise from the finite simulation space used in the Wigner-Weisskopf models. These errors are evident in simulations with one or more emitters, even in the limit of one excitation, in which we know that the HL model is exact. These errors are unavoidable, but do not significantly distort the predictions and decrease with increasing number of modes. There are additional errors that arise from the quality of the interpolation and the integration method, which can be improved in the near future. Finally, there are errors that arise from the theoretical predictions of the interaction constants~\eqref{Eq:GreenTensor}, which, as shown in other works~\cite{diaz-camacho2015}, do not exactly follow those laws at smaller than a wavelength separations. We expect that these errors will be improved by actually fitting the interaction constants in single-photon simulations and significantly decrease at long separations, the regime we study next.

\subsection{Two-level emitters in a quantum link}
\label{sec:quantum-link-benchmark}
In this section we will use the HL equations and the Wigner-Weisskopf model to explore a different limit, in which we have two emitters connected by a long waveguide. Our goal now is to explore the accuracy of the approximate models in a regime of very separated emitters that exchange photons with long retardations over many many periods. This is a particular limit of the previous model in which qubits decay way before their photons have reached the other node, $\gamma\tau_{12}\gg 1$.

We have performed simulations using qubits that are resonant with an even mode, $\phi_0=2\pi$, ensuring that $\gamma\tau_{12}=8\pi$ and that the probability of excitation of a qubit is negligible---around $e^{-8\pi}\sim 10^{-11}$-  --once the photon it created reaches the other end of the waveguide. Note that in this particular case the delay $\tau=L/v_\mathrm{g}$ fixes the length of the waveguide, removing one parameter from the Wigner-Weisskopf model. Unless stated otherwise, for the rest of the manuscript we fix the parameters of the bosonic waveguide in our physical model using the linear dispersion relation $\omega_k=v_g|k|$, $v_\mathrm{g}=1$ and we use $510$ photonic modes in our simulations with frequencies centered at the atomic transition $\omega_k=\Delta$. For the emitters, we assume the frequency transition of order of $ \Delta = 40 \gamma $, compatible with the strong coupling regime.

The outcome of these simulations are shown in Fig.~\SubFig{fig:two_qubit_closewavguide}. The first set of simulations in Fig.~\SubFig{fig:two_qubit_closewavguide}{a} starts with only one excited qubit, $\ket{\Psi(0)}=\ket{10}\ket{\mathrm{vac}}$, which relaxes, creating a photon that bounces back and forth in the waveguide, similar to the experiments in Ref.~\cite{zhong2018}. As before, at time $t=\tau_{12}$ the photon created by qubit 1 arrives at qubit 2, gets partially absorbed and partially reflected, and returns to qubit 1, which gets excited at $t=2\tau_{12}$. This sequence of alternating excitations is accurately reproduced by the HL model, with very small discrepancies, provided the time step is short enough and interpolation accurate. This is to be expected, since the single excitation limit is one in which the HL theory becomes exact (c.f. Sect.~\ref{sec:single-excitations}). More interestingly, we have performed the same simulations in the two-photon regime, exciting both emitters $\ket{\Psi(0)}=\ket{11}\ket{\mathrm{vac}}$ and allowing them to exchange excitations periodically at times $t=\tau_{12}, 2\tau_{12}, 3\tau_{12},$ etc. (c.f. Fig.~\SubFig{fig:two_qubit_closewavguide}{b-c}). These multiple collapses and revivals are accurately reproduced by both methods, further validating the utility of our theoretical framework and its application to explore new physics in the upcoming section.

\section{Cascaded super- and subradiant emission}
\label{sec:cascaded}
The previous benchmarks have shown evidence of collective phenomena in the spontaneous decay of the emitter array. This section will deepen our study of the physics of collective emission using as fundamental tool our delay differential HL equations, comparing the behavior of saturable emitters against linear systems, and exploring the emergence of collective dynamics both in the emitted light (i.e., superradiant bursts) and in the decay rates. While other studies show the emergence of atom-photon bound states with up to two excitations \cite{sinha2020,alvarez-giron2024} and supressed radiance has been observed in cascaded systems \cite{windt2025a}, the results in this section are novel and non achivable through any state-of-the-art method.

\subsection{Time-delayed collective emission}
\label{sec:superradiant-conditions}
The system under consideration is an array of $N$ equally spaced emitters in 1D free space, modeled with an infinite waveguide. We will focus on the resonant condition $\phi_0=k_0d=2\pi\times\mathbb{N}$, where emitted and received photons have similar phases. Under these conditions, the equivalent single-excitation Dicke subradiant and superradiant states are~\cite{dicke1954}
\begin{eqnarray}
	\ket{\psi_\mathrm{sub}} &= \frac{1}{\sqrt{N}} \sum_{m=1}^{N} (-1)^{m-1} \hat{s}_{m}^\dagger \ket{0}^{\otimes N}\ket{\mathrm{vac}} , \\
	\ket{\psi_\mathrm{sup}} &= \frac{1}{\sqrt{N}} \sum_{m=1}^{N} \hat{s}_{m}^\dagger \ket{0}^{\otimes N}\ket{\mathrm{vac}}.\nonumber
\end{eqnarray}
These states are called sub- and superradiant, respectively, because either they don't decay, $\ket{\psi_\mathrm{sub}}$, or because the rate of decay of the excitation is enhanced with respect to the dynamics of a single emitter, $\ket{\psi_\mathrm{sup}}$. These collective phenomena, which appear in the limit of zero delay $\tau=0$, can still be properly observed if we prepare the array in either of these states and study their dynamics, as we did in Sect.~\ref{sec:validation}.

\begin{figure}[t]
	\centering
	\includegraphics[width=0.7\linewidth]{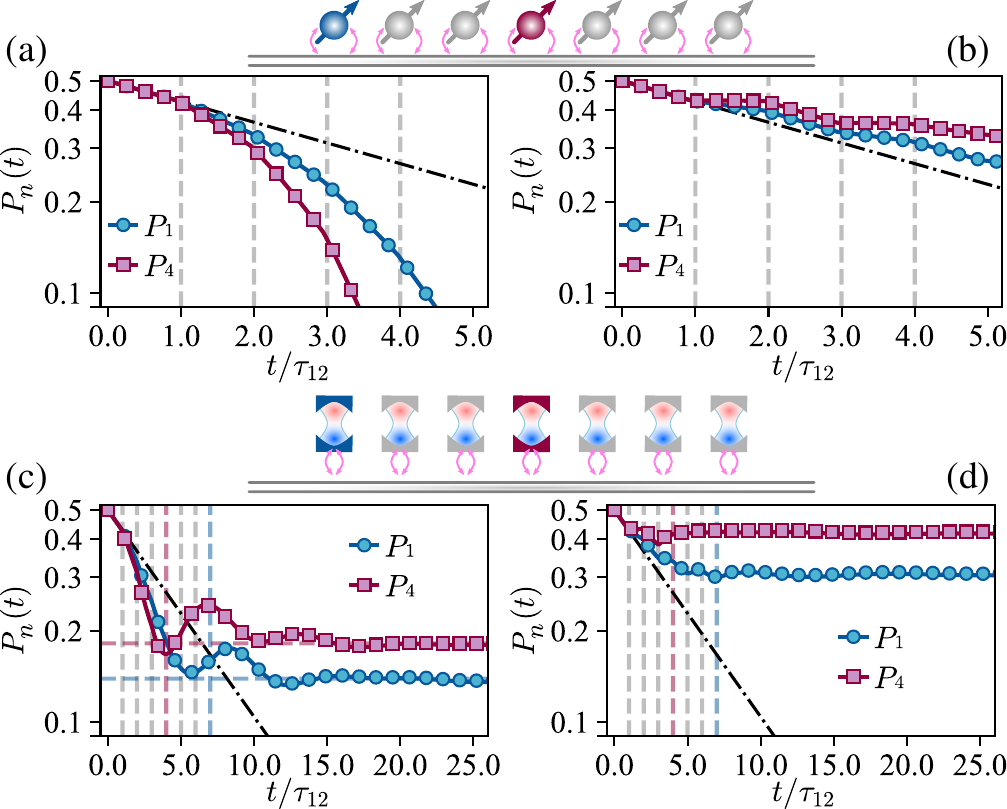}
	\caption{Population of the excited state $P_{n}(t)$ of the first and fourth emitter in a chain of 7 emitters. The blue lines represent the first emitter, while the purple ones represent the fourth emitter.  For all the  simulations we have fixed $\phi = 2\pi$   and $\gamma \tau_{0} = \pi / 20$. In Fig.(a),(b) the emitters are two-level emitter, and the solid line with dots represents the numerical integration of the HL equations. In figures (c), (d) the emitters are bosonic, and the solid lines are the analytical solution, while the dotted lines are the result of a numerical integrator that uses the same method that in the spin-like case. Plots (a), (c) correspond to a superradiant configuration with $ \ket{\psi (0)} = \ket{+++++++} $, while plots (b), (d) use a subradiant configuration with $\ket{\psi (0)} = \ket{+-+-+-+}$. In Fig.(c),(d) the solid line represents the analytical solution, and dot symbol represents the numerical integration of the HL equations}
	\label{fig:Super-Sub-Delay}
\end{figure}

However, the Dicke states are much harder to prepare in a non-local scenario than in a Markovian model with no retardations, where the emitters are close to each other. For this reason, our study will focus on product states that exhibit similar physics
\begin{eqnarray}
	\ket{\psi_{-}} &= \ket{+-+-+-\cdots} ,\\
	\ket{\psi_{+}} &= \ket{++++++\cdots},
\end{eqnarray}
with $\ket{\pm} = (1 + \hat{s}^\dagger)\ket{0}/\sqrt{2}$ the quantum superposition between 0 and 1 excitation. Note that the states $\ket{\psi_{+}}$ and $\ket{\psi_{-}}$ have projections over the states $\ket{\psi_\mathrm{sup}}$ and $\ket{\psi_\mathrm{sub}}$, respectively, and would also exhibit super- and subradiant behavior in the zero-delay limit.

In our first set of simulations, the array is prepared on either of these states, with the waveguide in vacuum $\ket{\Psi(0)} = \ket{\psi_\pm}\ket{\mathrm{vac}}$. These states are evolved using both the linear model of bosonic emitters, $\hat{s}_n(0)=\hat{b}_n$ and Eq.~\eqref{Eq:Jcoeff_infinity}, and the approximate equations for qubits, $\hat{s}_n(0)=\hat{\sigma}^-_n$ and Eq.~\eqref{Eq:sigmaInfiniteWG}. Our study initially focuses on the emitters' dynamics, computing the excitation probabilities $P_n(t)=\braket{\hat{s}_n^\dagger(t)\hat{s}_n(t)}$ as a function of time.

As an illustration of the resulting dynamics, Fig.~\ref{fig:Super-Sub-Delay} shows the decay process of a system with $N = 7$ emitters with commensurate phase $k_0d=2\pi$ and a small retardation $\gamma\tau_{12}=\pi/20$, plotting the probability of excitation of an outermost ($P_1$) and a central qubit ($P_4$). In both the qubit and in the bosonic systems, there are clear differences between the dynamics of the totally symmetric $\ket{\psi_{+}}$, Fig.~\SubFig{fig:Super-Sub-Delay}{a,c} and the alternating state $\ket{\psi(0)} = \ket{\psi_{-}}$, Fig.~\SubFig{fig:Super-Sub-Delay}{b,d}. The former states exhibit superradiant behavior, decaying faster than isolated qubits (dashed line), while the alternating states $\ket{\psi_{-}}$ exhibit subradiant behavior and, in the case of bosons, evidence of bound states in the continuum.

Other than a qualitatively similar enhancement and suppression of the decay, the dynamics is strikingly different from the Markovian master equation~\eqref{Eq:MasterEquation}. In the model without delays, all qubits experience the same dynamics due to the permutation invariance of the model and all-to-all interactions. In the one-dimensional retarded network, the delays become evident in the dynamics of the qubits, which experience sharp changes at times $t=\tau_{12}, 2\tau_{12}, 3\tau_{12}, $ etc., at which a qubit enters the light cone of yet another neighbor.

Let us focus in Fig.~\SubFig{fig:Super-Sub-Delay}{a}, inspecting the 1st and 4th emitters' dynamics. For times $0 < t \leq \tau_{12}$, neither of these qubits can receive any information from their neighbors. The dynamics of all emitters in this interval of time is governed by the exponential of a free emitter decay (black dashed line), with rate $\gamma$. During times $\tau_{12} < t < 2\tau_{12}$, the first emitter ($n=1$, the outermost qubit) is influenced by photons coming from one neighbor, while the central qubit ($n=4$) is influenced by two photons coming from qubits 3 and 5. The influence from those neighbors leads to an enhancement of the emission rate that is asymmetric, but in both cases the outcome is that the curves $P_{1}(t)$ (blue, circles) and $P_{4}(t)$ (purple, square) depart from the pure  exponential $\exp(-\gamma t)$ (dashed-dotted line). This enhancement of the decay is further accelerated as each emitter establishes connections with---or enters the light cone of---neighbors that are even further away.

This cascaded behavior and asymmetric distribution of time-dependent decay rates is not observed in a model such as a conventional master equation, where all emitters couple similarly (and locally) to the same modes. However, as in the master equation case, it remains the question of how much of this superradiant physics can be attributed to strong collective phenomena or merely to things such as a bosonic enhancement.

The first clue to this respect is found by comparing the bosonic and qubit simulations. These two models experience qualitatively similar behaviors, but with strong differences. For instance, the qubit decay more rapidly and to lower values than the bosons, which at some point reach a bound state. Furthermore, also in the subradiant states we find strong differences, the cavities quickly reaching an equilibrium state that is not seen in the same timescales for the qubits. The following subsections will deepen the study of these differences, looking for further qualitative and quantitative differences, both in the dynamics of the emitters as well as in the radiation produced by them.

\subsection{Time delayed superradiant burst}
\label{sec:Time-DelaySUPER}

In addition to the enhanced collective decay of certain quantum states, one of the phenomena observed in superradiant systems is the appearance of a collective superradiant peak in the output power generated by the array when all emitters are initially excited $\prod_n\hat{s}^\dagger_n\ket{0_n}^{\otimes N}$. More precisely, we look for abnormally high electromagnetic current flowing out of the system, which, for a short time, exceeds the current produced by the same number of independent emitters.

The output current of our system may be derived using input-output relations, integrating the field~\eqref{eq:field} and relating its dynamics to the state of the emitters. Generalizing the method by Gardiner and Collet to our non-Markovian model~\cite{gardiner1985} yields the following expression for the power emitted from the left of the chain. For the particular phase relation that we are exploring in this setup, the output intensity reads
\begin{eqnarray}
	& \langle \hat{I}_{\mathrm{out}}(t) \rangle   =  \langle \hat{a}^{\dag}_{\mathrm{out}}(t) \hat{a}_{\mathrm{out}}(t) \rangle \quad \mathrm{where } \label{eq:output_current}\\
	&  \hat{a}_\mathrm{out}(t) = -i\frac{\sqrt{\gamma}}{2} \sum_{n=1}^{N}\hat{s}_{n}^{-}(t-n\tau_{12}) \Theta(t-n\tau_{12}),\label{eq:output_field}
\end{eqnarray}
replacing $\hat{s}_n$ with $\hat{\sigma}^-_n$ or $\hat{b}_n$ depending on whether we look at qubits or linear emitters. The field leaving the array at one end of the chain, $\hat{a}_\mathrm{out}$ is a cascaded sum of the photons produced by each emitter, with the associated delays. This operator is formally identical to the retarded sums that drive the emitters' dynamics in Eq.~\eqref{Eq:sigmaInfiniteWG} or \eqref{Eq:Jcoeff_infinity}, because the output field of other qubits is what drives those emitters. Finally, note that, since we have an infinite waveguide, there should be another operator describing the output power at the other end, but because of symmetry considerations, and for the product states we work with, both operators lead to identical predictions, the power splitting equally on both sides.

\begin{figure}[t!]
	\centering
	\includegraphics[width=0.7\linewidth]{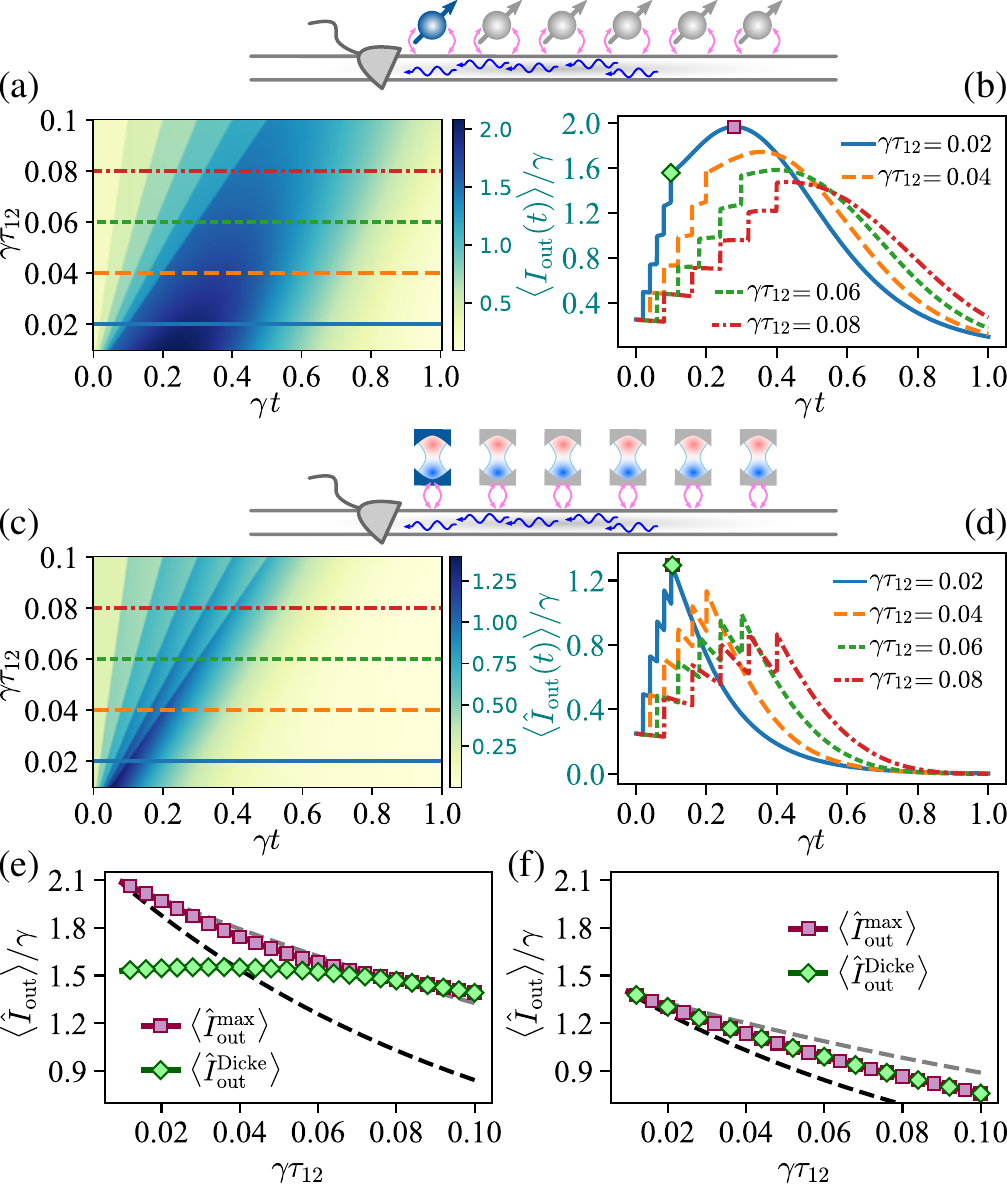}
	\caption{Output current $\langle \hat{I}_{\mathrm{out}}(t)  \rangle $ at the end of a chain of six emitters. We have fixed $\phi_{0} = 2 \pi $.  Subfigures (a),(c) show a heat map of the output current as a function of the time parameter $\gamma t$ and the separation between adjacent emitters, measured in $\gamma \tau_{0}$ for spin-like and bosonic emitters, respectively. Subfigures (b), (d) show the output current $\langle \hat{I}_{\mathrm{out}}(t)  \rangle $ for a set of fixed values of $\gamma \tau_{0}$ as a function of the dimensionless time $\gamma t $. The diamond and square dot symbols denote, respectively, the output currents $\langle \hat{I}_{\mathrm{out}}^{\mathrm{Dicke}} \rangle $ and $\langle \hat{I}_{\mathrm{out}}^{\mathrm{max}}\rangle $. In (e-f) we show the behavior of $\langle \hat{I}_{\mathrm{out}}^{\mathrm{Dicke}}  \rangle $ and $\langle \hat{I}_{\mathrm{out}}^{\mathrm{max}}  \rangle $ as a function of the time separation, for the two-level emitters case in (e), and the linear emitters case in (f). Dashed lines highlight exponential decay as $e^{-\gamma\tau_{12}/2}$ (gray) and $e^{-\gamma\tau_{12}}$ (black).}
	\label{fig:spin_heatmap}
\end{figure}

Figure \ref{fig:spin_heatmap} shows the power emitted by a chain of six emitters, initially prepared in the $\ket{\Psi(0)}=\ket{1}^{\otimes 6}\ket{\mathrm{vac}}$ state, as a function of time, $\gamma t$, and the separation between consecutive emitters, $\gamma \tau_{12}$. The output by both the cavities and the qubits exhibits discontinuities at times multiples of the separation between emitters, $t/\tau_{12}\in\{1,2,\ldots\}$, at which the light cone of two more qubits intersect. For 6 emitters, we observe a total of 5 discontinuous jumps, evidenced in 5 straight lines that reveal those light cones, both in Figs.~\ref{fig:spin_heatmap}a and Fig.~\ref{fig:spin_heatmap}c. The intensity radiated by the qubits and the cavities seem to exhibit peaks, but the behavior between light cones and the features of those peaks are markedly different.

First of all, note that in between light cones, while the power emitted by the qubits plateaus or slightly grows, for the linear emitters, the emission is further suppressed after every sharp raise. This can be clearly appreciated in Fig.~\ref{fig:spin_heatmap}d, especially in the curves corresponding to larger retardation values $\gamma \tau_{12}\sim 0.06,0.08$. In them, we can see that after every discontinuity the decay rate gets more negative. Unlike conventional systems, where the origin of slowdowns lays in dark states, in this system the slowdown is explained by bound photon excitations trapped by the quantum emitters~\cite{ordonez2006,tanaka2006,longhi2007,zhou2008,gonzalez-tudela2011,gonzalez-ballestero2013,facchi2016,facchi2018} that live in a superposition of being propagated among cavities, and in the  cavity modes themselves (c.f. Sect.~\ref{Sect:2LevelInf}).

Second and equally important, we find stark difference between the peaks exhibited by both types of arrays. For linear emitters, we appreciate an apparent ``enhancement'' of the intensity, which grows after each successive light cone intersection, but immediately after $t=5\tau$ the power decreases monotonically to zero. Indeed, in the limit of zero delay $\tau_{12}\to0$, the peak transforms into a pure exponential, revealing that we are only obseving the constructive interference of the light emitted by all cavities. In contrast, the qubits exhibit a more accelerated superradiant decay (c.f. Sect.~\ref{sec:superradiant-conditions}), reaching higher powers ($I=1.6$ vs. 1.5) at the threshold of full correlation $t= 5\tau_{12}$ when $\tau_{12} = 0.02$. This value, represented by the green diamond in ~\ref{fig:spin_heatmap}b, decreases as we increase $\gamma \tau_{12}$ . However, unlike the cavities, there is evidence in collective behavior in the continued increase of power, reaching higher values in a superradiant burst that persists for larger separations $\gamma\tau < 0.08$.

In Figs.~\ref{fig:spin_heatmap}e and~\ref{fig:spin_heatmap}f we show how the retardation time $\tau_{12}$ influences the output current and the emergence of the superradiant burst. To this end, we measure the output current at the exact time in which the system enters into the Dicke regime of the dynamics (all-to-all interactions), namely $\langle \hat{I}_{\mathrm{out}}^{\mathrm{Dicke}}  \rangle = \langle \hat{I}_{\mathrm{out}}(t= 5\tau_{12}) \rangle $, and the maximum output current $\langle \hat{I}_{\mathrm{out}}^{\mathrm{max}}  \rangle = \max_{t \geq 5\tau_{12}}\langle \hat{I}_{\mathrm{out}}(t) \rangle $ measured in the interval $t\geq 5\tau_{12}$, as highlighted by the square and circle dots in the Figs.~\ref{fig:spin_heatmap}b and~\ref{fig:spin_heatmap}d. It allows us to identify the superradiant burst happens in our dynamics when $\langle \hat{I}_{\mathrm{out}}^{\mathrm{max}}\rangle > \langle \hat{I}_{\mathrm{out}}^{\mathrm{Dicke}}\rangle$. It is possible to conclude that the values of maximum peaks exhibit an exponential decay with the retardation, with similar rates in the linear and qubit cases. For two-level emitters, superradiant burst appears immediately after the system enters the Dicke regime, except for long delay times, a limit in which the superradiant burst disappears and the emitted field intensity becomes comparable to that of independent emitters. In particular, for linear emitters there is no super-radiant burst, as we predict $\langle \hat{I}_{\mathrm{out}}^{\mathrm{max}}  \rangle = \langle \hat{I}_{\mathrm{out}}^{\mathrm{Dicke}}\rangle$ in this case (even in limit $\gamma\tau_{12} \rightarrow 0$). This is consistent with the fact that superradiance is a phenomenon mainly related to the nonlinear nature of the emitters~\cite{gross1982}.

\subsection{Maximal atomic emission rate scaling}
\label{sec:scaling}

As a last observable useful to witness collective effects, we study the time-delayed behavior of maximal atomic decay rate with respect to the number of emitters in the system~\cite{mok2024}. In the Heisenberg picture, the emission rate $R(t)$ is defined from the total atomic population operator $\hat{n}(t) = \sum_{j=1}^{N} \hat{s}_{j}^{\dagger}(t)\hat{s}_{j}$, according to
\begin{equation}
	R(t) = - \frac{d\braket{\hat{n}(t)}}{dt} ,
\end{equation}
with $\hat{s}_{j}(t) = \hat{\sigma}_{j}^{-}(t)$ and $\hat{s}_{j}(t) = \hat{b}_{j}(t)$ for two-level systems and cavity-like emitters, respectively.

Inspired by this definition, a second quantity of interest to our work is the \textit{logarithmic derivative} of the total population, defined as (in the Heisenberg picture)
\begin{equation}
	R_{\mathrm{ld}}(t) = -\frac{d}{dt} \log[\braket{\hat{n}(t)}] = -\frac{1}{\braket{\hat{n}(t)}}\frac{d\braket{\hat{n}(t)}}{dt} ,
\end{equation}
which also can be written as $R_{\mathrm{ld}}(t) = R(t) / \braket{\hat{n}(t)}$, interpreted as a population-normalized decay rate. The function $R_{\mathrm{ld}}(t)$ is particularly relevant to our discussion due to the time-delayed effects during the collective decay in our model. In fact, as already discussed from Fig.~\ref{fig:spin_heatmap}b), the bigger the delay time, the weaker the super-radiant burst observed by the emitted field. Therefore, through the quantity $R_{\mathrm{ld}}(t)$, we can compensate the effect of the population in the scaling even when the system is at the low-excitation limit, that is, the linear optics regime~\cite{gross1982, araujo2016}.

In general, the quantities depend on many different parameters, including the initial state of the system before the decay. Therefore, to find a scaling law for the quantities $R(t)$ and $R_{\mathrm{ld}}(t)$, with respect to $N$, one may consider the maximization~\cite{mok2024}
\begin{equation}
	R^{\star}(t,N) = \max_{\hat{\rho}(0)} R(t) , \quad R^{\star}_{\mathrm{ld}}(t,N) = \max_{\hat{\rho}(0)} R_{\mathrm{ld}}(t) , \label{Eq:emission_rates}
\end{equation}
over all possible initial states $\hat{\rho}(0)$ of the atomic emitters. In this way we can stablish bounds for the scaling law of our time-delayed non-Markovian dynamics. More precisely, considering the system of identically spaced emitters with nearest-neighbor time delay $\tau_{12}$, it is possible to show that from our generic HL equations we get
\begin{equation}
	R^{\star}(N,t) \leq \gamma\left( N + 2\sum_{n=1}^{N} (N-n) \Theta(t , n\tau_{12})  \right) .
\end{equation}
for the emission rate, and for the logarithm derivative emission rate we obtain
\begin{equation}
	R_{\mathrm{ld}}^{\star}(N,t) \leq \frac{\gamma}{N} \left( N + 2\sum_{n=1}^{N} (N-n) \Theta(t , n\tau_{12})  \right) ,
\end{equation}
which is valid for both two-level and cavity emitters. In particular, for cavity-like emitters, this scaling is valid up to a multiplication factor given by the excitation per emitter $N_{\mathrm{exc}}$.

\begin{figure}[t!]
	\centering
	\includegraphics[width=0.7\linewidth]{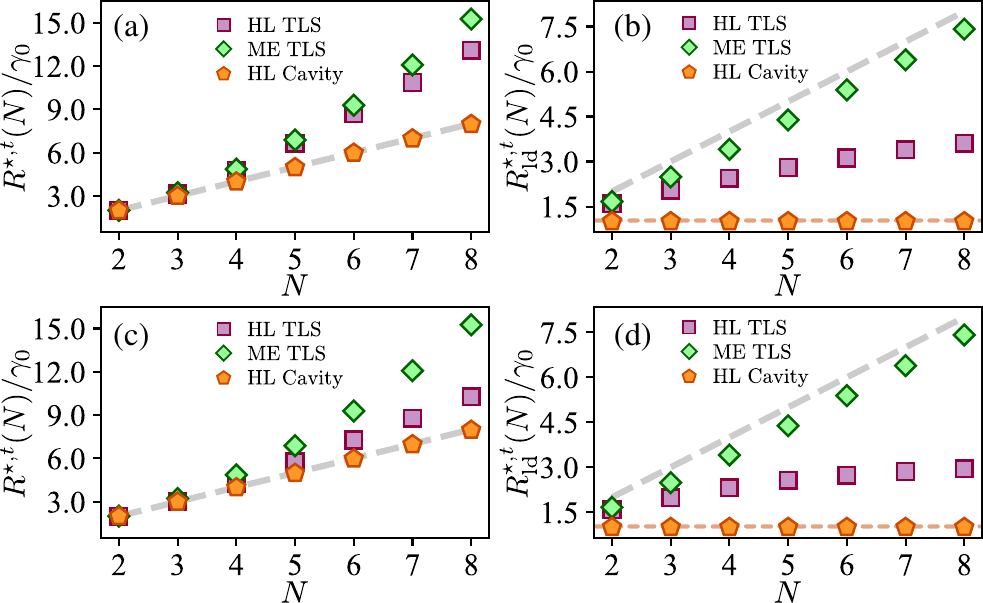}
	\caption{Scaling parameters $R^{\star}(N) / \gamma, R^{\star}_{ld}(N) / \gamma $ As a function of the number of emitters. The initial state is the fully excited state $\ket{\psi(0)} = \ket{1}^{\otimes N} $. Figures (a), (b) show the $\phi_{0} = \pi $ case, while Figures (c), (d) have $\phi = 2\pi$ }
	\label{fig:Scalings_N}
\end{figure}

It is possible to show that at the Markovian limit $\tau_{12} \rightarrow 0$, we obtain the upper bounds for the emission rate in the Dicke limit (all-to-all interactions) as $R^{\star}(N) \leq N^2$. In addition, according to our definition in Eq.~\eqref{Eq:emission_rates} we get $R_{\mathrm{ld}}^{\star}(N) = R^{\star}(N)/N \leq N$.

In addition to this upper bound, we investigate the scaling through numerical solution of our HL equations. To this end, we assume the decay process from the fully-excited state $\ket{\psi_{\mathrm{fe}}}$, as the system undergoes a super-radiant emission as shown in Fig.~\ref{fig:spin_heatmap}. In this case we are not maximizing the functions $R^{\star}(N,t)$ and $R_{\mathrm{ld}}^{\star}(N,t)$ over all possible states, instead we compute the maximal emission rate along the decay process as $R^{\star,t}(N) = \max_{t\geq 0} R(t)$ and $R^{\star,t}_{\mathrm{ld}}(N) = \max_{t\geq 0} R_{\mathrm{ld}}(t)$, with the initial state $\ket{\psi_{\mathrm{fe}}}$. The results are shown in Fig.~\ref{fig:Scalings_N}, where the scaling obtained by the HL equations for cavities and two-level systems, and by the master equations for two-level systems are provided for two different values of distances, namely $k_0 d = \pi$ and $k_0 d = 2\pi$ in Figs.~\ref{fig:Scalings_N}(a,b) and Figs.~\ref{fig:Scalings_N}(c,d), respectively.

As a first remark, in all cases considered in Fig.~\ref{fig:Scalings_N}, the scaling of the cavity-like emitters is linear, with $R^{\star,t}(N) = \gamma N $ and $R^{\star,t}_{\mathrm{ld}}(N) = \gamma$ what reinforces the previous discussion about the linear nature of the dynamics of such a system. For the two-level emitters case, we observe different behaviors for each distance considered. The master equations predictions provide the same behavior for $k_0 d = \pi$ and $k_0 d = 2\pi$, as in both cases we assume time instantaneous interactions and all-to-all collective behavior emerges from the beginning of the evolution. On the other hand, the HL equations capture the difference of the time-delayed all-to-all cooperativity for each case. For the set of parameters considered, the case $k_0 d = \pi$ provides a photon traveling time of $\gamma\tau_{12} = 0.04$, consequently for $k_0 d = 2\pi$ we find $\gamma\tau_{12} = 0.08$. In fact, the shorter the delay time, the closer the HL equations approximate the Master Equation results, as the all-to-all interactions start earlier during the decay process.

\section{Summary and discussion}%
\label{sec:discussion}
This work has addressed the study of quantum emitters with retarded interactions, a regime that is nowadays achievable in state-of-the-art photonic platforms~\cite{ferreira2024, storz2023, leung2019}. Our study has shown that, it is possible to to derive a set of non-Markovian operator equations that describe a combination of collective spontaneous emission and the correlated exchange of delayed photons travelling through the waveguide. These equations are exact for linear emitters, a situation in which the emitter's equations can be solved analytically and integrated numerically, for all initial states of the waveguide. Furthermore, we have provided evidence of an approximation technique that renders a similar delay-differential-equation that describes the dynamics of qubits in a waveguide that is initially empty. This is an approximate decoupling that is exact in the single-excitation limit, extremely accurate for two excitations and reproduces the limit of zero retardation from a typical master equation.

We have used the methods developed in this work to study the problem of collective emission in a low-dimensional environment. The study reveals a phenomenon that we term \textit{cascaded super- and sub-radiance.} This describes a dynamical behavior in which the emission of photons by qubits becomes increasingly correlated as the emitted photons are able to reach further apart neighbors. In this phenomenon, the finite speed of light and the light cones of the emitters become evident both in the emitter populations (c.f. Sect.~\ref{sec:superradiant-conditions}) and in the emergence of a superradiant burst in the generated light (c.f. Sect.~\ref{sec:Time-DelaySUPER}). This correlated physics is also evident in the enhancement of the emitter's decay, which significantly deviates from the linear behavior exhibit by cavities (c.f. Subsect.~\ref{sec:scaling}).

The study of this simple model has also revealed the persistence of subradiant states as the emitters separate from each other. In the regime in which the emitters are physically separated, these subradiant states are actually bound states in the continuum. However, in this setup those bound states include the possibility of photons bouncing back and forth among qubits, with a number of bound states that survive long separations and that increase with the number of emitters.

We believe that this study has demonstrated the interest of our approximate HL equations to understand non-Markovian networks of bosonic emitters, beyond the single-photon or chiral approximations These techniques are complementary to other methods such as the celebrated techniques based on tensor networks~\cite{vodenkova2024, feiguin2020, strathearn2018, pichler2016,grimsmo2015}. Indeed, it would be exciting, but outside the scope of this work, to benchmark the HL method with those TNs techniques, to understand the limits of our approximations. Another relevant avenue includes improving our DDE solvers to achieve greater accuracy and better interpolation, while simultaneously improving the speed. Finally, we expect that the DDEs should be amenable to other theoretical approximations, such as mean-field equations or truncated correlation expansions, that could provide analytical insight into the dynamical and asymptotic behavior of the emitter array.

Regardless of these improvements, we are confident that this line of research offers promising avenues also at the level of exploring and understanding the correlated physics that emerges in time non-local quantum many-body systems. At the lowest range of complexity remain the collective emission phenomena described in this work, which can be supplemented by studies of scattering and external drives. These 1D systems may also be modified to offer topological properties~\cite{ozawa2019}, either via SSH-like setups (alternating properties and interactions~\cite{bello2019}) or via the control of relative phases among nodes~\cite{hafezi2013, hafezi2011}. This last method may be also interesting for studying quasi-2D setups, such as one or two plaquettes, looking for frustrated physics. These complex phases and the competition between coherent and incoherent interactions is behind the emergence of photon bound states~\cite{ordonez2006, tanaka2006, longhi2007, zhou2008, gonzalez-tudela2011, gonzalez-ballestero2013, facchi2016, facchi2018, feiguin2020}, which may affect phenomena such as relaxation times and the thermalization of the emitter chain---both areas where already these small simulations exhibit interesting features.

\section*{Data availability statement}

The data and codes that support the findings of this study will be openly
available in GitHub following an embargo at the following URL: \href{https://zenodo.org/records/15634246}{https://zenodo.org/records/15634246}.

\section*{Conflict of interest}

There are no competing interests to declare.

\section*{Funding}

This work has been supported by Proyecto Sinérgico CAM 2020 Y2020/TCS-6545 (NanoQuCo-CM), the CSIC Research Platform on Quantum Technologies PTI-001 and from Spanish project PID2021-127968NB-I00. ACS is supported by the Comunidad de Madrid through the program Ayudas de Atracción de Talento Investigador ``César Nombela", under Grant No. 2024-T1/COM-31530 (Project SWiQL). JJGR acknowledges support by grant NSF PHY-2309135 to the Kavli Institute for Theoretical Physics (KITP). HJ gratefully acknowledges financial support from the program of China Scholarships Council (CSC202308620117).

\section*{Author contributions}

All authors have contributed equally to this work.

\section*{ORCID iDs} 

Carlos Barahona-Pascual\orcidlink{0009-0007-2372-9397} \href{https://orcid.org/0009-0007-2372-9397}{https://orcid.org/0009-0007-2372-9397}

\noindent Hong Jiang\orcidlink{0009-0008-7937-6671} \href{https://orcid.org/0009-0008-7937-6671}{https://orcid.org/0009-0008-7937-6671}

\noindent Alan C. Santos\orcidlink{0000-0002-6989-7958} \href{https://orcid.org/0000-0002-6989-7958}{https://orcid.org/0000-0002-6989-7958} 

\noindent Juan José García-Ripoll\orcidlink{0000-0001-8993-4624} \href{https://orcid.org/0000-0001-8993-4624}{https://orcid.org/0000-0001-8993-4624}

\section*{References} 
\bibliographystyle{iopart-num}
\bibliography{Waveguide-QED.bib}

\appendix

\section{Solution for linear emitters} \label{Sec:App:Solution}
In this section, we show how to compute the exact solution for the linear emitters. We remember that we seek for a solution of the form $ \hat{b}_{l}(t) = J_{lm}(t)\hat{b}_{m}(0) + $ (Noise terms) and that, for equally spaced emitters, the equations to solve are 
\begin{equation}
    \frac{d}{dt}J_{lm}(t) = \frac{-\gamma}{2}\sum_{n}e^{i\phi|l-n|}J_{nm}(t-\tau_{12}|l-n|) \Theta(t-\tau_{12}|l-n|) .
\end{equation}

This is a set of $N^{2}$ differential equations, consisting of $N$ subsets of $N$ coupled differential equations. If we take $J_{lm}(t) = J_{ml}(t)$, then each of these subsets contains all the functions for a single emitter. After applying this symmetry to the equations, the index $m$, that plays no role, can be removed, resulting in 
\begin{equation}
    \frac{d}{dt}J_{l}(t) = \frac{-\gamma}{2}\sum_{n}e^{i\phi|l-n|}J_{n}(t-\tau_{12}|l-n|) \Theta(t-\tau_{12}|l-n|).
\end{equation}

In order to avoid exponential behavior, we use the variable change $J_{l}(t) = e^{\frac{-\gamma t }{2}}f_{l}(t)$ and get 
\begin{equation}
    \frac{d}{dt}f_{l}(t) = \frac{-\gamma}{2}\sum_{n\neq l} e^{\left( i\phi + \frac{\gamma \tau_{12}}{2} \right) |l-n|}f_{n}(t-\tau_{12}|l-n|) \Theta(t-\tau_{12}|l-n|). \label{eq_app:derivartive}
\end{equation}

These equations are continuous in intervals of duration $\tau_{12}$, and each $\dot{f}_{l}(t)$ has terms that activate once $t > |l-n|\tau_{12}$ for every $n$

Since $J_{lm}(t<0) =0 $ and $  J_{lm}(0) = \delta_{l,m} $, the solution in the interval $t \in [0,\tau_{12})$ becomes, trivially,
\begin{equation}
    f_{l}(t) = \delta_{l,m} \Theta(t). \label{eq_app:solution_0}
\end{equation}

After this, the derivative of each $f_{l}(t)$ is also determined in the interval $[\tau_{12},2\tau_{12}) $, according to Eq. ~\eqref{eq_app:derivartive}. Thus, we can plug Eq. ~\eqref{eq_app:solution_0} into  Eq. ~\eqref{eq_app:derivartive} and integrate it with limits of integration $\int_{\tau_{12}}^{t}$ to ensure continuity, to extend the solution to $t\in[0,2\tau_{12})$. This results in 
\begin{equation}
    f_{l}(t) = \left[\delta_{l,m}\right]\Theta(t) +\left[(\delta_{l,m+1}+\delta_{l,m-1}) \frac{-\gamma}{2}e^{i\phi + \frac{\gamma \tau_{12}}{2}}(t-\tau_{12}) \right] \Theta(t-\tau_{12}).\label{eq_app:solution_1}
\end{equation}

This is, of course, assuming that both $f_{m\pm1}(t)$ exist (if we only have two qubits this is not the case, for example). The step function $\Theta$ arises because this contribution only appear when $t>\tau_{12}$.

This process can be iterated to extend the solution to $t \in [0,(m+1)\tau_{12})$ for any $m$ desired. In each step, for each function $f_{l}(t)$, you collect the terms that correspond to each delay $|l-n|$ and that you have not taken into account yet,  integrate them over $\int_{m\tau_{12}}^{t} $ and add $\Theta(t-m\tau_{12}) $. The resulting term will modify each of the other $f_{l}(t)$ in the future only once.  This results in a solution of the form 
\begin{equation}
    f_{l}(t) = \sum_{j}e^{j(i\phi + \frac{\gamma \tau_{12}}{2})}\left[\sum_{k=0}^{j}\mathcal{C}_{lm,jk} (t-j\tau_{12})^{k} \right]\Theta(t-j\tau_{12})
\end{equation}
It is precisely these coefficients $\mathcal{C}_{lm,jk}$ that we calculate with the algorithm described in this appendix. 

As a final remark, the result of this calculation is in agreement with  a numerical integrator similar to the one used for two-level emitters.

\section{Master equation and HL correspondence \label{Sec:appendix_b}}

Let us consider a generic master equation of the form
\begin{equation}
\partial_t \rho = -i[H, \rho] - \sum_{n,m}\frac{\Gamma_{nm}}{2}(2A_n \rho B_m - B_m A_n \rho - \rho B_m A_n).
\end{equation}
It can be shown~\cite{Gardiner2014} that the expectation value of any observable $\hat{O}$ is equivalent to that of a Heisenberg-picture version of it, $\braket{\hat{O}}_t := \mathrm{tr}(\hat{O}\rho(t)) = \mathrm{tr}(\hat{O}^H(t)\rho(0))$, with the operator $\hat{O}^H(t)$ that satisfies $\hat{O}^H(0) = \hat{O}$ and
\begin{equation}
	\partial_t\langle\hat{O}^H\rangle = \braket{\left (-i[\hat{O}^H, H] - \sum_{n,m}\frac{\Gamma_{nm}}{2}
	[B_m, \hat{O}^H]A_n + B_m [\hat{O}^H, A_n]\right )}.
\end{equation}

Following the master equation for our model with $N$ emitters (29) we conclude that the coherences of the emitters satisfy the set of equations
\begin{equation}
	\partial_t \langle \hat{\sigma}^{-}_{n}  \rangle   = \sum_{l,j} \frac{\Gamma_{lj}}{2} \left( \langle [\hat{\sigma}^{+}_{j},\hat{\sigma}^{-}_{n}] \hat{\sigma}^{-}_{l} \rangle + \langle \hat{\sigma}^{+}_{j}[\hat{\sigma}^{-}_{n},\hat{\sigma}^{-}_{l}]\rangle\right)  -i \sum_{l,j} \Delta_{l,j}^{LS} \langle [\hat{\sigma}^{-}_{n},\hat{\sigma}^{+}_{l}\hat{\sigma}^{-}_{j} ] \rangle . 
\end{equation}
Using now $ [\hat{\sigma}^{+}_{l}, \hat{\sigma}^{-}_{n}] = \hat{\sigma}^{z}_{l} \delta_{l,n}$ along with $[\hat{\sigma}^{-}_{n}, \hat{\sigma}^{-}_{j}] = 0 $, This reduces to
\begin{equation}
    \partial_t \langle \hat{\sigma}^{-}_{n} (t) \rangle   = \sum_{l} \left( \frac{\Gamma_{ln}}{2} +i  \Delta_{l,n}^{LS} \right)  \langle \hat{\sigma}^{z}_{n}(t) \hat{\sigma}^{-}_{l}(t) \rangle  =   \sum_{l} \frac{\sqrt{\gamma_{n}\gamma_{l}}}{2}e^{ik_{0}|x_{l}-x_{n}|}  \langle \hat{\sigma}^{z}_{n}(t) \hat{\sigma}^{-}_{l}(t) \rangle . 
\end{equation}
which correspond to the limit of zero retardation from our time-delayed differential equations~\eqref{Eq:sigmaInfiniteWG} . 

\end{document}